\documentclass[prb,twocolumn,epsf,epsfig,amsmath]{revtex4}
\usepackage[toc,page]{appendix}
\usepackage{graphicx,pdflscape}% Include figure files
\usepackage{color}
\usepackage{bm}% bold math
\usepackage{alltt,dsfont}
\usepackage{appendix}
\usepackage{amsmath,amssymb}
\newcommand{\llabel}[1]{  \label{#1} }

\newcommand {\apgt} {\ {\raise-.5ex\hbox{$\buildrel>\over\sim$}}\ }
\newcommand {\aplt} {\ {\raise-.5ex\hbox{$\buildrel<\over\sim$}}\ }
\newcommand{\lessim}{\aplt}

\newcommand{\iden}{ \mathds{ 1}}
\newcommand{\G}{{\cal{G}}}
\newcommand{\GH}{{\bf g}}
\newcommand{\GHI}{\GH^{-1}}
\newcommand{\bh}[1]{{\, \tilde{#1}}}
\newcommand{\tr}{{\text {Tr} }}
\newcommand{\V}{{\mathcal V}}

\newcommand{\X}[2]{X_{{#1}}^{#2}}
\newcommand{\si}{\sigma}
\newcommand{\sib}{\bar{\sigma}}
\newcommand{\tJ}{\ $t$-$J$ \ }
\newcommand{\beq}{\begin{eqnarray}}
\newcommand{\eeq}{\end{eqnarray}}
\newcommand{\barray}{\begin{eqnarray}}
\newcommand{\earray}{\end{eqnarray}}
\newcommand{\nn}{\nonumber}
\newcommand{\A}{\hat{{\mathcal A}}}
\newcommand{\Ag}{{\mathcal A}}
\renewcommand{\AA}{{{\mathcal A}}}
\newcommand{\llv}{\lVert}
\newcommand{\rrv}{\rVert}
\renewcommand{\lll}{\langle \langle}
\newcommand{\rrr}{\rangle \rangle}
\newcommand{\disp}[1]{Eq.~(\ref{#1})}
\newcommand{\dispop}[1]{~{\bf (\ref{#1})}}
\newcommand{\refdisp}[1]{Ref.~(\onlinecite{#1})}

\newcommand{\Ham}{{{\cal H}}}
\newcommand{\Hamc}{\hat{{\cal H}}}
\newcommand{\up}{\uparrow}
\newcommand{\dn}{\downarrow}

\newcommand{\I}{{\cal I}}

\newcommand{\chem}{{\bm \mu}}
\newcommand{\ch}[1]{{C}_{#1}}
\newcommand{\chd}[1]{{C}^\dagger_{#1}}
\newcommand{\n}[1]{ {N}_{#1}}
\newcommand{\chl}[1]{\widetilde{C}_{#1}}
\newcommand{\chdl}[1]{\widetilde{C}^\dagger_{#1}}
\newcommand{\muin}{(\mu^{-1})}

\newcommand{\vv}{\nu}
\renewcommand{\k}{~  \hat{k}\, v_f}

\begin{document}
\title{   Theory of   extreme correlations   using canonical Fermions and path  integrals    }
\author{ B Sriram Shastry  }
\affiliation{Physics Department, University of California,  Santa Cruz, Ca 95064 }
\date{25 February 2014}
\begin{abstract}
The \tJ model is studied using a novel and rigorous   mapping of the Gutzwiller projected electrons, in terms of canonical electrons. The mapping  has  considerable similarity to the Dyson-Maleev transformation relating  spin operators to canonical Bosons. This representation gives rise to a non Hermitean quantum theory, characterized by  minimal redundancies. A path integral representation of the canonical theory is given. Using it,   the salient results of the  extremely correlated Fermi liquid (ECFL)  theory, including the  previously found Schwinger  equations of motion,   are easily rederived. Further  a transparent physical interpretation of the previously introduced auxiliary Greens functions and the ``caparison factor'' is obtained.

The low energy  electron spectral function in this theory, with a strong intrinsic  asymmetry,  is summarized  in terms of a  few expansion coefficients. These include an important  emergent energy scale $\Delta_0$ that shrinks to zero   on approaching  the insulating state,
thereby making it difficult to access the underlying very low energy Fermi liquid behavior. 
 The scaled low frequency ECFL spectral function,    related simply  to the Fano line shape, has  a  peculiar energy dependence unlike that of a Lorentzian. The resulting energy dispersion obtained by maximization  is  a  hybrid of a massive and a  massless  Dirac spectrum   $
 E^*_Q\sim  \gamma\, Q- \sqrt{\Gamma_0^2 + Q^2} $, where the vanishing of $Q$, a momentum type variable, locates the kink momentum. Therefore the  quasiparticle velocity  interpolates between $( \gamma \mp 1 )$ over a width $\Gamma_0$ on the two sides of $Q=0$,     implying  a  kink there that  resembles a prominent  low energy feature  seen in   angle resolved photoemission spectra (ARPES) of  cuprate materials. We also propose novel ways of  analyzing the ARPES data to isolate the predicted asymmetry between particle and hole excitations.
\end{abstract}
\maketitle

%\tableofcontents
%Notes; Rationalized the definition of the Anamolous Gorkov functions

\section{Introduction}

 The intensely studied  \tJ model is often regarded
  as the effective low energy Hamiltonian for
   describing several  observed  phenomena  in cuprate superconductors \cite{Anderson}.  Here the  $U \to \infty$ limit is presupposed, and hence the Hilbert space is  restricted  to  a maximum of single occupancy at each site, i.e.  Gutzwiller projected\cite{Gutzwiller}.   A few words on the choice of the \tJ  model are relevant here. The implied infinite $U$ limit   eliminates  high energy ($U$ scale) electronic states,  known as the  upper Hubbard band states.   The  residual low energy ($ \lessim 100 $ meV scale) excitations are  probed by sensitive spectroscopies and transport phenomena,  making the \tJ model suitable for our task.  At reasonably high $U$, say comparable to the band width in  a Hubbard model, this elimination of the upper Hubbard band must already occur in part.  Therefore the  limit  $U\to \infty$ must be regarded as a useful mathematical idealization of the very strong, or extreme correlation phenomenon. The resulting Gutzwiller projected electron operators,  denoted by  Hubbard's convenient notation of $X$ operators \cite{Hubbard}, are rendered non canonical.   The  non-canonical nature of the electrons precludes the  Wick's theorem  underlying  the Feynman diagram approach, whereby leading to the fundamental difficulty of the \tJ model, namely  the impossibility  of  a straightforward  Feynman type perturbative expansion.  This situation leads to enormous  calculational difficulties, so that  systematic and controlled analytical calculations with this model have been very difficult. 
 
In a series of recent papers \cite{ECFL,Monster,Hansen-Shastry,Asymmetry,Anatomy,DMFT-ECFL,Large-D}, we have shown that it is possible to overcome some of these difficulties by using alternate methods  based  on Schwinger's treatment of field theory with time dependent potentials. This idea  yields exact equations of motion for the electron Greens function.   These equations have the nature of functional differential equations, and provide a powerful launching pad for various approximations. The specific approximation pursued  is a systematic expansions in a parameter $\lambda$ related to double occupancy. Using this we have  presented an analytical  theory of the normal state of the \tJ model termed the  extremely correlated Fermi liquid  (ECFL) theory. An   interesting feature of   the theory  is that  we find a non-Dysonian representation of the projected  electron Greens function. This is  a significant    structural  departure  from the usual field theories, and  arises in a  most natural fashion. The  Greens function  is determined by  a {\em  pair} of self energies, denoted by  $\Phi(\vec{k}, i\omega_n)$ and $\Psi(\vec{k}, i\omega_n)$, instead of the standard  Dyson self energy $\Sigma(\vec{k}, i\omega_n)$ (see \disp{twin-self} below).  The latter can be reconstructed from the pair by a simple inversion.  Starting with rather simple  pairs of self energies,    it is found that non trivial complexity is introduced into the Dyson self energy through this inversion process.
Explicit self consistent  calculations in parameter $\lambda$ have been carried out to $O(\lambda^2)$ so far, and yield reliable results for electron densities $0 \leq n \lessim .7$. The  detailed dynamical results of the ECFL theory have been benchmarked against independent theories in overlapping domains; e.g. against high temperature series results  in \refdisp{High-T}.  The ECFL theory has been shown to have a momentum independent Dyson self energy in  the limit of infinite dimensions \refdisp{Large-D}. This enables benchmarking  against the dynamical mean field theory (DMFT)  in \refdisp{DMFT-ECFL}.   Importantly,  the results from the ECFL theory for the spectral function compare well with a large $U$ Hubbard model   solved by the DMFT method, and not just infinite $U$. 
 The ECFL theory has also been benchmarked in \refdisp{AIM} against the exact solution  of the
asymmetric $U = \infty$ Anderson impurity model, obtained from the numerical renormalization group study  of  Krishnamurthy, Wilson  and Wilkins \refdisp{KWW}.  In addition, a detailed comparison between the data on cuprate superconductors at optimal filling and the theoretical  photoemission spectral lines of the ECFL theory has been carried out  in \refdisp{Gweon-Shastry} and \refdisp{Kazue-Gweon}, where excellent agreement is found.
In all cases  studied,  the comparisons with  ECFL  are good,   and seem to indicate the  utility of this approach.

 The ECFL formalism   could  initially seem   somewhat unfamiliar, in view of  its reliance  on the analysis of the Schwinger equations of motion. This analysis  was  originally used  to derive  the main constituents of the theory, namely the auxiliary Greens function and the two self energies  (detailed  below).  This type of analysis is somewhat  removed  from the  toolkit of ``standard'' many body physics courses, and hence might obstruct a ready visualization of these objects. One goal of the present work is to show that these results are (A)  minimal, i.e.  having least redundancy, and (B)  available  more transparently. The latter follows  from an important and novel {\em hat removal theorem}, leading to a  compact mapping of the Hubbard operators to canonical Fermions. The
 mapping  is given in  \disp{rephats22}  and  described further in Section~(\ref{unhermitean}),   leading  to a  path integral formulation (Section~(\ref{pathintegral})).  It is possible that such a  simplified presentation could  lead to improved strategies for devising approximate methods, especially close to the insulating state. 

This method  rests on an exact replacement rule for the Hubbard $X$ operators in terms of the canonical Fermi operators 
\barray
\X{i}{0 \si} \to \ch{i \si}, \; \X{i}{ \si 0} \to  \chd{i \si}(1- N_{i \sib}) ,\;\X{i}{\si \si'}\to\chd{i \si} \ch{i \si'}. \llabel{rephats22}
\earray
This replacement rule is shown to be exact when  ``right-operating''  on states which satisfy the Gutzwiller constraint.  This  replacement  is similar   in spirit to the Dyson-Maleev representation \refdisp{Dyson}, \refdisp{Maleev}, where  spin operators are expressed in terms of canonical Bosonic operators. With the advantage of this representation, most steps in the ECFL theory, such as the factorization of the Greens function into an auxiliary Greens function, the two self energies and the caparison function (see Eqs~( \ref{caparison}, \ref{auxg}, \ref{twin-self}))  becomes very intuitive.

The analogy  can be  pushed further to establish a parallel between the $\lambda$ parameter of the ECFL theory, and the  small parameter  of the Dyson Maleev \cite{Dyson,Maleev} theory, namely the inverse spin $\frac{1}{2 s}$. Finally we are able to make contact with the  illuminating  work of Harris, Kumar, Halperin and Hohenberg \refdisp{HKHH}. In a detailed work these authors computed the Greens function of the spins for  two sublattice antiferromagnet using the Dyson-Maleev scheme and extracted the lifetime of the magnons of the theory. We find that their calculation contains the  precise Bosonic counterparts of the auxiliary Greens function and the  second self energy $\Psi$ defining the ``caparison function'' of the ECFL theory (see Eqs~( \ref{caparison}, \ref{auxg}, \ref{twin-self})).   Unlike the spin problem with variable number of excitations, the \tJ model has a fixed number of particles. Hence there are significant new elements in the ECFL theory involving  the imposition of the Luttinger Ward volume theorem, as discussed later.

A few comments on the  canonical description of the equations of motion are appropriate. The general problem is to represent a time evolution of a state of the \tJ model 
\beq
[\psi]'_{final} = Q'_M\ldots Q'_2.Q'_1 .  [\psi]'_{initial}, \label{eq330}
\eeq
where the primed states and operators are in the \tJ model Hilbert space defined with the three allowed states at each site as usual (see Sec ~(\ref{preliminaries}) for details). The operators $Q'_j$ may be thought of as the exponential of the \tJ Hamiltonian: $Q'_j \sim e^{- i t_j H_{tJ}}$ written in terms of the projected operators. Since the algebra of the projected  electrons  is very inconvenient, one seeks a reframing of the problem into a canonical space. This involves   mapping the states,  the  Hamiltonian and all other operators of the original theory, into the unconstrained Hilbert space of two Fermions at each site. This canonical space is of course described by the usual Fermi operators $C_{j \up}, C_{j \dn}$ and their  adjoints. This gives us an enlarged space with four states per site, with one redundant state corresponding to double occupancy, eliminated using
 Gutzwiller projection. There are  various possibilities for doing this elimination leading to the different theories in literature. This includes the popular  slave Boson or slave Fermion technique \cite{slave,Read,lee}, where  additional degrees of freedom, over and above the already enlarged 4 dimensional local state space, are introduced and finally eliminated as best as possible. This handling of the redundancy  leads to gauge theories for the \tJ model that are reviewed in \refdisp{lee}.

In the enlarged state space 
 let us block diagonalize  the state space into physical and unphysical states  and write the projection operator as
\beq
[\psi]= \begin{bmatrix} \psi^{ph} \\ \psi^{un} \end{bmatrix}; \;\; 
\hat{P}_G = \begin{bmatrix} \iden^{ph} &0 \\ 0 &0 \end{bmatrix},
\eeq
where $\iden^{ph}$ is the identity operator in the physical space. The relevant operators in  the theory $Q_j$  e.g. the Hamiltonian, the creation operators or the destruction operators, are now written in terms of the canonical Fermions: 
\beq
Q_j =\begin{bmatrix} Q_j^{p p} & Q_j^{p u} \\
Q_j^{u p} & Q_j^{u u}
\end{bmatrix}. \label{eq32}
 \eeq
The next goal of the construction is to ensure that a state resulting from the application of a sequence of operators  on a projected state remains in the projected space, i.e.
\beq
[\psi]_{final} = Q_M\ldots Q_2.Q_1 . \hat{P}_G . [\psi]_{initial}, \label{eq33}
\eeq
and $[\psi]_{final}=\hat{P}_G.[\psi]_{final}$. If this condition is not ensured, the projector has to be introduced at all intermediate time slices, thus making the calculations intractable.
  A sufficiency condition for this is the commutation $[Q_j,\hat{P}_G]=0$ for all $j$. The slave Boson- Fermion technique    uses the conservation of the Gutzwiller  constraint by writing a suitable version of the  Hamiltonian. This enables the use of  a time independent Lagrange multiplier, as demonstrated   in the   work of Read and Newns \refdisp{Read}.  In  Sec~(\ref{hermitean}), we display a compact  Hermitean representation that also achieves this, without however adding further states (beyond the four states) into the problem.

   While the commutation condition $[Q_j,\hat{P}_G]=0$ is sufficient, it is  not necessary, and a much  less restrictive condition can be found. We note that if the operators $Q_j$ have a vanishing $Q_j^{up}$ then the product in \disp{eq33} remains in the physical sector with
 \beq
[\psi]_{final} = \begin{bmatrix} Q^{pp}_M\ldots Q^{pp}_2.Q^{pp}_1. ~   \psi^{ph}_{initial} \\0\end{bmatrix} .
\label{eq34}
\eeq
 The property of a commuting projection operator $[Q_j, \hat{P}_G]=0$,  requires  that $Q_j^{pu}=0$ as well as $Q_j^{up}=0$, whereas the  vanishing property of the unphysical components noted in \disp{eq34} requires only  $Q_j^{up}=0$.  Then $Q_j^{pu}$  as well as $Q_j^{uu}$ are quite arbitrary. 
  With this property, all the $Q_j$ operators in \disp{eq32} are block triangular 
 \beq
Q_j =\begin{bmatrix} Q_j^{p p} & Q_j^{p u} \\
0 & Q_j^{u u}
\end{bmatrix}. \label{eq322}
 \eeq
 In more formal terms the  sufficiency condition with least constraints that leads to  \disp{eq34} (via  the block triangularity \disp{eq322})  is
 \beq
 (\iden - \hat{P}_G) . Q_j . \hat{P}_G =0. \label{eq36}
 \eeq
 This condition is  also expressible as  $ [Q_j, \Hat{P}_G].\Hat{P}_G=0$;   a conditional vanishing of the commutator, when  right operating on projected states. This observation  provides some intuition for why \disp{eq36} is sufficient in the present context.
 In view of the block triangular operators in \disp{eq322}, the adjoint property, namely
 of representing conjugate operators by their  matrix  Hermitean conjugates, 
  is lost in this representation. This is   seen clearly  in \disp{rephats22}, where the first two operators are mutual adjoints in the defining representation, but not so in the canonical basis. In general this  situation is expected to lead to non Hermitean Hamiltonians.
 The non Hermitean representation in \disp{rephats} and Sec~(\ref{unhermitean}) implements  this idea and therefore leads to the most efficient  canonical theory. We show  that it exactly matches the minimal theory, found from the minimal description of the \tJ model in terms of the Hubbard $X$ operators and the Schwinger equations of motion.  It is  notable that the Gutzwiller projection operator does not appear  {\em explicitly}  in the equations of motion, although it does play a crucial role in the canonical  theory, and is at the root of its difficulty.

The plan of the paper is as follows. In Sec~(\ref{preliminaries}, \ref{schwinger-eom}, \ref{second-chem}) we review the Schwinger equations of motion for the \tJ model, and the ingredients  of the  recent method developed for a systematic expansion in a parameter $\lambda$. In Sec~(\ref{2D}) we summarize the general form of the Greens function at low frequencies near the Fermi surface, and obtain the prototypical spectral function of the theory. We summarize in Sec~(\ref{2E}) a   kink in the electronic dispersion that arises from the theory, and seems to be  closely related to that seen in many photoemission experiments. We also present simple but important ideas for analyzing photoemission data, with a view to isolating important feature of asymmetry predicted by the ECFL theory.

In Sec~(\ref{canonical-fermions}) we formulate the ``best possible'' representation of the Hubbard operators in terms of  canonical Fermions, as discussed above.    
Sec~(\ref{hermitean}) summarizes the well known representation and Sec~(\ref{unhermitean})
implements the block triangular idea to obtain a non Hermitean method with least redundancy.
Sec~(\ref{3C}, \ref{KMS}) give further details of the Hamiltonian in this representation and the proof of the antiperiodic temporal boundary conditions necessary for defining the new framework. 

In Sec~(\ref{product-ansatz}), the above non Hermitean  representation is  used to analyze the nature of the  Greens function of projected electrons. Quite remarkably  this process also  yields  the Greens function as a convolution of an auxiliary Greens function and a caparison function, in complete parallel to that obtained from the Schwinger method employed in Sec~( \ref{schwinger-eom}, \ref{second-chem}).  In Sec~(\ref{lambda-Fermions}) we generalize the above representation to define $\lambda$ Fermions where the Gutzwiller projection is only partial, and becomes full at $\lambda=1$. The equations of motion from these Fermions are shown to be those obtained in the $\lambda$ expansion of Sec~(\ref{second-chem}).

In Sec~(\ref{Dyson-Maleev}) we display a close analogy between the non Hermitean representation of the Gutzwiller projected electrons and the well known Dyson-Maleev representation of spin operators in terms of canonical Bosons. This connection also provides further   meaning of the small parameter $\lambda$ in the Fermion theory, as a parallel of the expansion parameter $\frac{1}{2s}$ of the Dyson Maleev theory.  A connection with the  work of Harris, Kumar, Halperin and Hohenberg (HKHH) \cite{HKHH} is noted, who invented a method for computing the lifetime of spin waves in antiferromagnets, with  considerable overlap with our representation of the Greens function with two self energies. 

In Sec~(\ref{pathintegral}), we cast the canonical theory in terms of Fermionic path integrals, and show how the exact Schwinger equations of motion can be obtained directly from this representation,  thereby validating  all the links in the argument.
The subtle role of the Gutzwiller projection operator is explored, it does not appear explicitly in the equations of motion and yet plays an important role in the theory. In Sec~(\ref{conclusions}) we summarize the main points of the paper.

In Appendix~( \ref{Minimal}) we summarize the derivation of the minimal equations of motion from the Schwinger viewpoint. In Appendices~(\ref{coherent-states}, \ref{pathint}, \ref{pauli}) we  provide the details of the coherent state path integrals and the implementation of the Gutzwiller projection. In Appendix~(\ref{AE}) we provide a more detailed interpretation of the 
caparison function in terms of a change of variable of the source fields.

\section{Summary of the ECFL theory for the \tJ model}
\begin{table*}[t]
\begin{center}
\begin{tabular}{ p {.6 in}  p{ .6in} p{1.25 in} p {.6 in} p {1.25 in} p {1.25 in} p {1.25 in} }\hline 
Step(I)& Step(II)& Step(III)& Step(IV)& Step(V)& Step(VI)& Step(VII) \\ \hline
 Green's function $\G$ in terms of Hubbard operators&Exact Schwinger equations of motion for $\G$.&  Product expression into canonical part $\GH$ and adaptive spectral weight (caparison) part $\mu(k)$. &  Exact  equations for $\GH(k)$ and $\mu(k)$. & Introduction of interpolating parameter $\lambda$ connecting the Fermi gas to the extreme correlation limit. &   
Shift invariance  requires  second chemical potential $u_0$. Same sum rule for both GreenÕs functions so that Fermi surface volume is conserved.&   Successive orders in $\lambda$ expansion satisfying shift invariance for practical calculations. \\  
    $\G$ &$\partial_\tau \G$&$\G(k)= \GH(k) \mu(k)$&&$0 \leq \lambda \leq 1$&$\sum \G = \sum \GH = \frac{n}{2}$&  \\ \hline
\end{tabular} 
\caption{ A flowchart of the ECFL theory as developed in \refdisp{ECFL} and \refdisp{Monster}.  See Sections~(\ref{schwinger-eom}, \ref{second-chem}) for a detailed description.   \label{flowchart}}
\end{center}
\end{table*}

\subsection{The \tJ model preliminaries \label{preliminaries} }
  The well studied  \tJ model is a two component Fermi system on a lattice, defined on the restricted subspace of  three  local states, obtained by excluding all doubly occupied configurations.  The allowed states are  $|a\rangle$ with $a=0,\up,\dn$, and the double occupancy state $|\up \dn\rangle$ is removed by the (Gutzwiller) projection operator. 
These Gutzwiller projected electron operators are denoted, in  the convenient notation due to Hubbard, as $\X{i}{a,b} = |a\rangle \langle b|$. 
Its Hamiltonian $H_{tJ}$ is expressed in terms of the $X$ operators so that the single occupancy constraint is explicit. Summing over repeated spin indices we write
\barray
H_{tJ}&=& H_t+H_J, \nn \\
H_{t}&=& - \sum_{i j } t_{ij} \X{i}{\si 0}\X{j}{0\si} - \chem \sum_i \X{i}{\si \si},  \nn \\
H_J&=&  \frac{1}{2} \sum_{ij} J_{ij} \left( \vec{S}_i . \vec{S}_j - \frac{1}{4} \X{i}{\si \si} \X{j}{\si' \si'}   \right). \label{hamiltonian}
\earray
In computing the Green's functions we  add  two kinds of   Schwinger  sources to the Hamiltonian;  the anticommuting Grassman pair $J, J^*$ coupling to electron creation and destruction operators,  and the commuting potential $ \V$,  coupling to the   charge as well as spin  density.  These sources  serve to generate compact Schwinger equations of motion (EOM),  and are set to zero at the end.  Explicitly we write
\barray
\A_S & = & \sum_i \int_0^\beta   \A_S(i, \tau) d \tau, \nn \\
\A_S(i, \tau) &=&  \left[  \X{i}{\si 0}(\tau) J_{i \si}(\tau) + J^*_{i \si}(\tau) \X{i}{ 0 \si}(\tau)\right] \nn \\
&&+ \V_{i}^{\si' \si}(\tau) \X{i}{\si' \si}(\tau), \llabel{sources}
\earray
and all time dependences are as in $Q(\tau)=e^{\tau H_{tJ}}Q e^{-\tau H_{tJ}}$.  The generating functional of Green's functions of the \tJ model is
\beq
Z[J,J^*,\V]\equiv \tr_{tJ} \ e^{-\beta H_{tJ}}  T_\tau \left( e^{- \A_S}  \right). \llabel{part1}
\eeq
it reduces to the standard partition function on turning off the indicated source terms. The Green's functions for positive  times $0 \leq \tau_j \leq \beta$,
 are defined  through the Martin-Schwinger  prescription \cite{MS,angleaverage}:
\beq
\G_{\si \si'}(i  \tau_i, f  \tau_f) = - \frac{ \langle T_\tau \left( e^{- \A_S}   \X{i}{0 \si}(\tau_i) \X{f}{\si' 0}(\tau_f)\right)\rangle}{   \langle T_\tau e^{- \A_S} \rangle} \ .  \label{gdef}
\eeq
The functional $Z$ conveniently yields  the Green's functions upon
 taking  functional derivatives with respect to the sources, e.g.
\beq
\G_{\si \si'}(i \tau_i, f \tau_f)  =\left( \frac{1}{Z} \frac{\delta^2 Z}{ \delta J^*_{i \si}(\tau_i) \delta J_{f \si'}(\tau_f)}\right),  \llabel{green}
\eeq
where the sources are turned off at then end.  We note that $n_\si$,  the number  of particles per site, is determined from the number sum rule:
\beq
n_{\si}= \G_{\si \si}(i \tau^-, i \tau), \label{number-sumrule}
\eeq 
and $\chem$ the chemical potential is fixed by this constraint.

\subsection{The Schwinger equations of motion \label{schwinger-eom}}
The detailed theory of the \tJ model developed so far \cite{ECFL,Monster} uses the Schwinger equations of motion. Since these equations play a fundamental role in the theory, we summarize next  the equations of motion and their extension, obtained by introducing a parameter $\lambda$. We  relegate to  Appendix ~(\ref{Minimal})  the  derivation of the ``minimal theory'' equations. In the minimal theory,
the most compact set of Schwinger equations are established, and some redundant terms from \refdisp{ECFL} are omitted. 
This minimal version of the theory is important for the purposes of the present paper, since our goal in this paper is to recover these from a canonical formalism.

 As the Schwinger school  has\cite{MS,Kadanoff-Baym,KM}  emphasized, a field theory is rigorously determined by its equations of motion  plus the boundary conditions.  We can also  establish alternate descriptions such as path integrals formulations, from the requirement that they reproduce these   equations of motion- we present an example of this approach  in Section~(\ref{path2}). In terms of the original description of the \tJ model involving the Hubbard $X$ operators, the Schwinger equation of motion is a partial differential equation in time and also a functional differential equation  involving the derivatives with respect to a Bosonic source:
\beq
&&  \left(  \GHI_{0, \si_i,  \si_j}(i \tau_i,  j \tau_j) -   \hat{X}_{ \si_i \si_j}(i \tau_i, j \tau_j)-    {Y_1}_{ \si_i \si_j}(i \tau_i, j \tau_j)\right) \nn \\
 &&\times   \G_{\si_j \si_f}( j \tau_j, f \tau_f)  = \delta_{if} \delta(\tau_i-\tau_f)  \left( \delta_{\si_i \si_f} - \gamma_{\si_i \si_f }(i \tau_i) \right),  \nn \\ \label{Minimal-eq}
 \eeq
 where $\GH_0$ is the  noninteracting Green's function \disp{gnon}, $\hat{X}$ is a functional derivative operator \disp{xopdef}, $\gamma$ is the local Green's function obtained from $\G$
as $\gamma_{\si_a \si_b}(i \tau_i) = \si_a \si_b \G_{\sib_b \sib_a}(i \tau_i^-, i \tau_i)$ (see \disp{gamma-def})
  and  $Y_1$ is the band hopping times $\gamma$ \disp{ydef};  further details can be found in the Appendix~(\ref{Minimal}).   This  equation  has been written down in \refdisp{ECFL} and \refdisp{Monster}:
 Antiperiodic boundary conditions with respect to both times (as in Eqs.~(\ref{27a}) and (\ref{27b})),  and the number sum-rule \disp{number-sumrule} together with the equation of motion \disp{Minimal-eq}, specify  the theory  completely.

\subsection{ The $\lambda$ expansion,  the shift identities and second chemical potential $u_0$  \label{second-chem} }

The  idea of introducing a parameter into the EOM \disp{Minimal-eq} becomes quite natural when we observe the Schwinger EOM for  the Hubbard model closely. These can be written schematically as $\left(  \GHI_{0} -  U \delta/{\delta{\V}} -  U  G \right). G = \delta \   \iden $. 
  By scaling the interaction $U \to \lambda \ U$, with a parameter $\lambda$ ($0\leq \lambda \leq 1$),  the interacting theory is connected continuously to the Fermi gas by tuning $\lambda$ from $1$ to $0$.
The standard  perturbative expansion  can be organized by  counting   the various powers of $\lambda$,   setting $\lambda=1$ at the end before evaluating the expressions\cite{comment-perturb}. Below in Section ~(\ref{lambda-Fermions}) we provide a more microscopic argument for introducing the $\lambda$ parameter in the Hubbard $X$ operators directly, this method leads back to the equations found here.

 In the corresponding equation for the  \tJ model \dispop{Minimal-eq}, we observe  that the 
Green's function differs from that for the free Fermi gas through two terms on the left hand side, exactly as in  the Hubbard model, but also   through one term on the right hand side.  Scaling these three terms by $\lambda$, we rewrite \dispop{Minimal-eq} schematically  as: 
\beq\left(  \GHI_{0} -  \lambda \hat{X}-  \lambda {Y_1}\right). ~\G = \delta \  ( \iden  - \lambda {\gamma}  ). \label{Min-2}
\eeq
The strategy of the perturbative $\lambda$ expansion method is to 
build up the solution of this equation at $\lambda=1$ through a suitable  expansion  in $\lambda$, starting from the free Fermi limit $\lambda=0$. Thus $\lambda < 1$ corresponds to the admixture of a finite fraction of double occupancy that vanishes at $\lambda=1$. Insights from sum rules, the skeleton graph expansion  and the physics of the Hubbard sub bands has played a major role in formulating a systematic $\lambda$ expansion described in detail in \refdisp{ECFL} and  \refdisp{Monster}.

Within this approach  it is also necessary to add a term
$\lambda u_0 \sum_i N_{i \up} N_{i \dn}$ to the Hamiltonian,  and a corresponding term to the EOM, so that the $ \hat{X}$ and $Y_1$ in \disp{Min-2} are suitably redefined. Here
  $u_0$ is an extra Hubbard interaction type parameter that is  {\em determined  by a sum rule} as explained below.  At $\lambda=1$ such a term makes no difference since the double occupancy is excluded.
   This parameter $u_0$ also enables us to enforce a simple but crucial symmetry of the \tJ model- {\em the shift invariance}, noted   in \refdisp{Monster}. This invariance  arises from the twofold function  of the hopping in the \tJ model when expressed in terms of the canonical operators,  of providing hopping as well as the four Fermion (interaction) terms. Therefore in an exact treatment, adding a constant times the identity matrix to  the  hopping matrix: $t_{ij} \to t_{ij}+ \mbox{const}\times \delta_{ij}$,  shifts the center of gravity band innocuously.  In   approximate implementations it has  the unphysical effect  of  also adding to  the interaction (i.e. four Fermion type) terms.  Such a change   must therefore  be compensated by an adjustable parameter that can soak up this additive constant. Indeed   $u_0$ provides precisely this type of a parameter.  It also plays the role  of a second chemical potential $u_0$ (\refdisp{Monster}) to fix the number of Fermions in the auxiliary Green's function $\GH$ through $n_{\si}= \GH_{\si \si}(i \tau^-, i \tau)$, while the thermodynamical  chemical potential $\chem$ (residing in the non interacting $\GHI_0$), is fixed by the number sum rule $n_{\si}= \G_{\si \si}(i \tau^-, i \tau) ($\disp{number-sumrule}).
Enforcing this shift invariance  to each order in  the $\lambda$ expansion plays an important ``watchdog'' role on the $\lambda$ expansion, in addition to other standard constraints such as the Ward identities.

   To summarize some key points of the $\lambda$ expansion, we first decompose the Greens function into the space time convolution of an auxiliary Greens function and a caparison function as:
  \beq
\G= \GH . \mu. \label{factor2}
\eeq
With this the operator in  equation \dispop{Min-2} acts on the two factors of \disp{factor2}, and  breaks  into two equations upon using  the {\em ansatz} that $\GH$ has a canonical structure $\left(  \GHI_{0} -  \lambda \hat{X}-  \lambda {Y_1}\right). \GH = \delta \   \iden  $. The $\lambda$ expansion \refdisp{Monster}  is then an iteration scheme that  proceeds by an expansion of  the caparison function $\mu(k)$ and  $Y_1$ ($Y_1= t \gamma$) in powers of $\lambda$.  Dyson's skeleton graph idea is implemented by keeping the auxiliary $\GH$ intact ( i.e. unexpanded in $\lambda$), while all other variables are expanded in powers of $\lambda$ {\em and} $\GH$, thereby obtaining self consistent equations for $\GH$ and the vertex functions.
 Successive levels of approximation are obtained by retaining increasing powers of $\lambda$.  At each approximation level,  we set $\lambda=1$ before actually evaluating the expressions, and  implement 
 the  antiperiodic boundary conditions \dispop{27a}, \dispop{27b}, and  the number sum-rule $n_{\si}= \G_{\si \si}(i \tau^-, i \tau)$ (\disp{number-sumrule}).
 
Elaborating on the  representation \disp{factor2} of $\G$, we note that 
the  $\gamma$ term on the right hand side of \dispop{Min-2} is due to the non canonical anticommutator of the projected Fermi operators.  As noted in \refdisp{ECFL}, this term contains the essential difficulty of the \tJ problem, having no parallel in the (canonical) Hubbard type models.   After turning off the sources, in the momentum-frequency space we can further  introducing two self energies $\Psi(k, i\omega)$, and $\Phi(k, i\omega)$
with
\beq
\mu(\vec{k}, i \omega_n)&=& 1- \frac{n}{2} + \Psi(\vec{k}, i \omega_n) \label{caparison} \\
\GH^{-1}(\vec{k}, i \omega_n)&=& \GH_{0}^{(-1)}(\vec{k}, i\omega) - \Phi(\vec{k}, i \omega_n), \label{auxg}
\eeq
where the constant $\frac{n}{2}$ in \disp{caparison} is fixed by the condition that $\Psi$ vanishes at infinite frequency. The auxiliary Greens function satisfies a second sumrule analogous to \disp{number-sumrule}, written in the Fourier domain: 
\beq
(k_B T) \sum_{k, n} e^{ i \omega_n 0^+} \GH_{\si \si}(k, i \omega_n) = n_\si \label{second-sumrule}.
\eeq
Clearly the same sumrule  holds for $ \G_{\si \si}(k, i \omega_n)$.  \disp{factor2} can now be written explicitly  in the non-Dysonian form proposed in \refdisp{ECFL} and  \refdisp{Anatomy}
 \beq
\G(\vec{k}, i\omega)= \frac{1- \frac{n}{2} + \Psi(\vec{k}, i\omega)}{\GH_{0}^{(-1)}(\vec{k}, i\omega)- \Phi(\vec{k}, i\omega)}. \label{twin-self}
\eeq
As argued in \cite{ECFL,Monster,Asymmetry,DMFT-ECFL}, simple Fermi liquid type self energies $\Psi$ and $\Phi$ can, in the combination above,   lead to highly asymmetric (in frequency) Dyson self energies from the structure of \disp{twin-self}, thus providing a considerable tactical advantage in describing extreme correlations.
We further discuss     the physical meaning of this decomposition and the twin self energies  in Section~(\ref{product-ansatz}).  Table ~(\ref{flowchart}) provides an overview of the various steps in the construction of the theory.

\subsection{  $\G(\vec{k}, i \omega_n)$  and the low energy spectral function  in ECFL theory \label{2D}}
We summarize here the low temperature low energy theory near the Fermi surface that follows from the general structure of \disp{twin-self} in terms  of a small number of   parameters,  upon assuming that the two self energies have  a Fermi liquid behavior at low energies.  In the limit of large dimensions, a similar exercise gives a very interesting spectral function that matches the exact solution of the $U= \infty$ Hubbard model found from the dynamical mean field theory (DMFT) \refdisp{DMFT-ECFL}. The presentation below generalizes that to include a momentum dependence that is absent in high dimensions, and is supplemented by a discussion of the behavior of the various coefficients as the density of electrons $n$ approaches unity, or equivalently the hole density $\delta \to 0$.

The Dyson self energy can be inferred from a simple inversion, and has  a strong set of corrections to the Fermi liquid theory that we delineate here. We assume here a Fermi liquid type state that survives the limit of small hold density $\delta \to 0$.  In reality at very small $\delta$ several other broken symmetry states would compete and presumably win over the liquid state, so that this Fermi liquid state would be metastable. It characteristics are of interest and hence we proceed to describe these.

We study \disp{twin-self} by analytically continuing $i\omega \to \omega+ i 0^+$ and write 
\beq
\GH_{0}^{(-1)}(\vec{k}, i\omega) = \omega + \chem - (1-\frac{n}{2} ) \varepsilon_k 
\eeq
Let us define $\hat{k}$ as the {\em normal deviation} from the Fermi surface i.e. $\hat{k}= (\vec{k}- \vec{k}_F). \vec{k}_F/|\vec{k}_F|$, and  the frequently occurring Fermi liquid function
\beq
{\cal R}= \pi \{ \omega^2+ (\pi k_B T)^2\}. \label{rdef}
\eeq
We carry out a low frequency expansion  as follows:
\beq
1- \frac{n}{2} +\Psi(\vec{k},\omega) &=& \alpha_0 + c_\Psi (\omega + \vv_\Psi \k) \nn \\
&& + i {\cal R}/\gamma_\Psi + O(\omega^3), \label{psiex}
\eeq
where $\alpha_0= 1-\frac{n}{2}+ \Psi_0$  is the constant term at the Fermi surface,
and a similar expansion for $\Phi(\vec{k},\omega)$ so that 
\beq
 &&\omega +\chem - (1-\frac{n}{2} ) \varepsilon_k - \Phi(k,\omega)  = \nn \\
&&  (1+ c_\Phi ) \left(  \omega - \vv_\Phi \k + i {\cal R} /\Omega_\Phi +O(\omega^3) \right), \label{phiex}
\eeq
where $v_f =(\partial_k \varepsilon_k)_{k_F}$ is the {\em bare} Fermi velocity.  The expansion coefficients above are in principle functions of the location of $\vec{k}_F$ on the Fermi surface, and have suitable dimensions to ensure that $\Psi$ is dimensionless and $\Phi$ is an energy. The  dimensionless  velocity renormalization constants $\vv_\Phi$ and $\vv_\Psi$  capture  the momentum dependence normal to the Fermi surface, arising  from   the two respective self energies.
 The Greens function near the Fermi surface can now be written as
\beq
\G(\vec{k}, \omega) \sim \frac{z_0}{\alpha_0} \left(\frac{\alpha_0 + c_\Psi ( \omega + \vv_\Psi \k) + i {\cal R}/\gamma_\Psi}{\omega- \vv_\Phi \k + i {\cal R}/\Omega_\Phi}\right)
\eeq
where $z_0= \alpha_0/(1+c_\Phi)$ is the net quasiparticle renormalization constant. The spectral function can be computed from $A(\vec{k},\omega) = - \frac{1}{\pi} \Im m ~ \G(\vec{k}, \omega+ i 0^+)$ in the ECFL form of a  Fermi liquid function times a  caparison function $\mu(k,\omega)$ as follows:
\beq
A(\vec{k},\omega)= \frac{z_0}{\pi} \frac{\Gamma_0}{(\omega- \vv_\Phi \k)^2 + \Gamma_0^2} \times \mu(k, \omega) \label{spectral-function},
\eeq
where the (Fermi liquid)  width function (or decay rate)
\beq
\Gamma_0( \hat{k}, \omega)= \eta + \frac{ \pi (\omega^2+ (\pi k_B T)^2)}{\Omega_\Phi}, \label{width}
\eeq
with an extra  phenomenological  parameter $\eta$ required  to describe elastic scattering \cite{Gweon-Shastry} in impure systems. The  caparison function is
\beq
\mu( \hat{k}, \omega) = 1- \frac{\omega}{\Delta_0} +  \frac{\vv_0 \k}{\Delta_0}, \label{caparison2}
\eeq
 where we introduced an important (emergent) low energy scale combining  the other parameters:
 \beq
 \Delta_0 = \alpha_0 \frac{\gamma_\Psi}{\Omega_\Phi - c_\Psi \gamma_\Psi} \label{delta}
 \eeq
 and the dimensionless momentum dependence coefficient
  \beq
  \vv_0= (\vv_\Psi  \gamma_\Psi c_\Psi+ \vv_\Phi \Omega_\Phi)/( \Omega_\Phi - c_\Psi \gamma_\Psi). 
  \eeq 
   A  cutoff  $\theta\left(\mu( \hat{k}, \omega)\right)$  is implicit in \disp{caparison2}, so that the function $\mu(\hat{k}, \omega)$  is assumed to be zero at large positive frequencies as discussed in \refdisp{ECFL}.
  The five final   parameters defining the spectral function \disp{spectral-function}  are $z_0, \vv_0,\vv_\Phi, \Omega_\Phi, \Delta_0$.  For  fitting experimental data,  it may be best to   think of them as adjustable parameters  that determine the line shapes, their asymmetries  and also features in the spectral dispersions. In addition the $\eta$ parameter is needed to describe impurities that are not contained in the microscopic theory.  In the early fit \cite{Gweon-Shastry} the total  number of free parameters is even smaller-just two instead of five. The corrections to the Landau Fermi liquid theory are encapsulated in the caparison factor, which contains a correction term that is odd in frequency and seems to be ultimately responsible for the asymmetric  appearance of the line shapes \cite{Gweon-Shastry,Asymmetry}.

  For reference we note that  in the limit of high dimensions \cite{DMFT-ECFL}, the coefficient  of the momentum dependent term $\vv_0$ vanishes in \disp{spectral-function}, while the earlier fits to experiments in \cite{Gweon-Shastry}, it is non zero, and in modified fits \cite{Kazue-Gweon} its magnitude is varied to get a good description of  the constant energy cuts of the data.

It is useful to consider the approach to the Mott insulating limit, where the parameters behave in a specific fashion to satisfy the expected behavior. We consider the limit of density $\delta \to 0$, and a frequency scale $ 0 \leq |\omega| < \omega_c \sim \delta t$, where the above expression \disp{spectral-function} may be expected to work.    For reference, it is useful to note that in this limiting case,  the widely used Gutzwiller-Brinkman-Rice theory \cite{Gutzwiller, BR} gives  the quasiparticle propagator as: 
\beq
 G_{GBR}
 (\vec{k},\omega) \sim \frac{z}{\omega- z \k},  \label{br22}
 \eeq
where $z$ vanishes linearly with $\delta$ as $z= 2 \delta/(1+\delta)$. This leads to a delta function spectral weight $A_{GBR}= z \,\delta(\omega - z\k)$. In contrast  \disp{spectral-function} provides the spectral function at non zero $T$ and $\omega$.

As $n \to 1$ in \disp{psiex}  we expect that the constant   $\Psi_0 \to - \frac{n}{2}$, in order  to reach the Mott insulating limit continuously. This implies that 
$\alpha_0 \propto \delta$ in this regime,  and this drives the various other coefficients as well. We summarize the expected behavior of  the above five coefficients
\beq
z_0 &\to& \overline{z}_0  \times \delta \nn \\
\Delta_0 &\to& \overline{\Delta}_0 \times \delta \nn \\
\Omega_\Phi & \to & \overline{\Omega}_\Phi \times \delta \nn \\
\vv_0 & \to&  \overline{\vv}_0 \times \delta \nn \\
\vv_\Phi & \to&  \overline{\vv}_\Phi \times \delta 
\label{limiting} 
\eeq
 by using  an overline for denoting a non vanishing limit of the stated variable \cite{DMFT-ECFL, 2DMFT}. The scaling of the velocity constants $\vv$ is guided by the results in high dimensions, and ensure that the dispersing quasiparticles have a vanishing bandwidth as we approach the insulator- as emphasized by Brinkman and Rice \cite{BR}.
  From this we find  that the  ECFL spectral function \disp{spectral-function} satisfies a simple homogeneity  (i.e. scaling) relation valid in the low energy regime for a scale parameter $s$:
  \beq
 A(  \hat{k}, s \,\omega | s \,T,  s \, \delta) = A(\hat{k}, \omega|  T , \delta), \label{scaling}
 \eeq
where the dependence on the temperature and hole density are made explicit. The momentum variable does not scale with $s$ due to the assumed behavior of the $\vv$'s.   The scaling  holds for $\eta=0$, and generalizes to a non zero values  if we scale  $\eta \to s\, \eta $.
This scaling relation describes a Fermi liquid including  significant corrections to Fermi liquid theory through the caparison function. It rests upon   the  specific behavior for the coefficients as the density varies near the insulating state, unlike other generalized scaling relations that  have been proposed in literature \refdisp{senthil} for non Fermi liquid states. 
If set $s \times \delta = \delta_0$ with say $\delta_0 \lessim .5 $, then  the ratio $\frac{\delta_0}{\delta} \gg 1$ and we infer 
\beq
A(\hat{k}, \omega|  T , \delta) \sim A(\hat{k} , \omega \frac{ \delta_0}{ \delta} | T \frac{  \delta_0}{\delta} , \delta_0), \label{scaling2}
\eeq relating the low hole density system to an overdoped  (i.e.  high hole density) system at a high effective temperature. This  relation provides basic  intuition for  why the \tJ model,  near the insulating limit  behaves  almost like a  classical  liquid,  unless one fine tunes parameters  very close to the $T=0, \omega=0$ limit.

\subsection{   Electronic origin of the low energy kink   and further tests  of dynamical asymmetry \label{2E}}
In this section we summarize the origin of the  important low energy {\bf kink}  feature of the dispersion relation obtained in the ECFL theory. 
Since a similar feature is seen in the experiments on angle resolved photoemission studies (ARPES) of various groups \cite{Johnson,Lanzara,Campuzzano,Gweon-Shastry}, it is worth clarifying the purely electronic origin of this feature within the ECFL theory.  A higher (binding) energy kink is also seen and is well  understood in terms of the behavior of the  self energy over a greater range \cite{Anatomy,DMFT-ECFL}, and is not pursued here. Rather we focus on the low energy kink seen around $-.05$ eV in several compounds \cite{Johnson,Lanzara,Campuzzano,Gweon-Shastry},  and  finds a  natural interpretation within ECFL. 

 We also present a few experimentally testable features relating to {\bf dynamical asymmetry}, i.e.  the asymmetric in $\omega$ correction to the Fermi liquid theory contained in ECFL, arising from the caparison function in \disp{spectral-function}.

Let us assume that $|\omega| \ll \Gamma_0$ at low enough frequency relative to $T$ so that we may treat $\Gamma_0$ as a constant.
We may then bring \disp{spectral-function} to an interesting  form  studied in \refdisp{Anatomy} by defining variables
\beq
\epsilon& =& \frac{\omega- \vv_\Phi \k}{\Gamma_0} \nn \\
\sinh{u_k} & =& \frac{\Delta_0 + (  \vv_0- \vv_\Phi) \k }{\Gamma_0} \label{sinh-variable},
\eeq
so that the spectral function reduces to  the  standard form occurring in the ECFL theory:
\beq
A(u_k ,\epsilon) = A_0 \frac{\sinh u_k - \epsilon}{1+ \epsilon^2} \times \theta(\sinh u_k - \epsilon) \label{scaled-spectrum}
\eeq
 with $A_0= \frac{z_0}{\Delta_0}$.
This expression is valid for small enough $\epsilon$\cite{Anatomy,ECFL},
and   can be viewed as the (weighted) sum of the real and imaginary parts of a simple damped  oscillator with a  scaled susceptibility $\chi(\epsilon)= 1/(\epsilon + i )$.
  It is interesting to note that  the scaled spectral function \disp{scaled-spectrum} can be related to  the
(scaled)  Fano line shape  
  \beq
  A_{Fano}(q_f,\epsilon) \propto \frac{(q_f+ \epsilon)^2}{(1+ \epsilon^2)}. \label{Fano}
  \eeq
This spectrum   is  often considered with the Fano parameter $q_f>0$, it is  highlighted by  a vanishing  at negative energies  $\epsilon = - q_f$, representing the destructive interference of a scattering amplitude with a background term arising from a continuum of states. However we can  flip the sign of $q_f$ and 
  by  choosing $q_f = - e^{u_k}$,  we can relate these through 
  \beq
  A({u_k} ,\epsilon)\propto \left( A_{Fano}(-e^{u_k}, \epsilon)- A_{Fano}(  -e^{u_k}, \infty) \right). \nn \\ \label{connect-Fano}
  \eeq
For the purpose of representing ARPES spectral functions,   the scaled spectral function  \disp{scaled-spectrum} gains an advantage over the Fano line shape \disp{Fano}   by the absence of a background  at large $|\epsilon|$. In relating them via \disp{connect-Fano},  the background term in the   Fano process is killed, while its  interference with the peak is retained.
      
%%%%    
Unlike the simple Lorentzian  obtained at $u_k \to \infty$, the energy variable enters the numerator as well as the denominator in  both \disp{scaled-spectrum} and the Fano shape. This feature gives rise to  the characteristic  skew to the ECFL spectrum. The spectral function can be maximized with respect to the frequency at a fixed $\hat{k}$, 
yielding the energy distribution curve (EDC)  dispersion $E^*_k$, or with respect to  $\hat{k}$ at a fixed frequency $\omega$,  giving  the momentum distribution curve (MDC) dispersion $E_k$.  Let us introduce  the convenient variables
\beq 
r= \frac{\vv_0}{\vv_\Phi},
\eeq
giving the ratio of the two velocity factors. The ratio $r=0$   in the limit of high dimensions \cite{DMFT-ECFL}.  In the simplified ECFL analysis in \cite{ECFL, Gweon-Shastry}, we find $r>1$  due to the suppression of $\vv_\Phi$ relative to $\vv_0$ by a quasiparticle renormalization factor $z_{FL}$. We see below that the magnitude and sign of  $(r-1)$  play a significant role in determining the location of the kink, and its observability in ARPES respectively. We also introduce a (linear in $\k$)  energy variable:  
 \beq Q(\hat{k})= \Delta_0 + (\vv_0-\vv_\Phi) \k.  \label{Qdef} \eeq 
 In terms of these,  the two dispersions are obtained as  
 \beq
 E(k)&=& \frac{1}{2-r} \left(  \vv_\Phi \k + \Delta_0 - \sqrt{r (2-r) \, \Gamma_0^2  + Q^2} \right), \nn \\
 && \label{MDC} \\
 E^*(k)&=&  \left(  \vv_0 \k + \Delta_0 - \sqrt{\Gamma_0^2+ Q^2} \right). \label{EDC}
 \eeq
Simplifying the notation,  both energy  dispersions are of the form $E\sim \gamma \,Q- \sqrt{Q^2+M^2}$, i.e. the hybrid of a massless and a massive Dirac spectrum. As $Q$ varies from $-\infty$ to $\infty$, the energy crosses over from $(\gamma+1)Q$ to $(\gamma-1)Q$, thus exhibiting a knee or a kink near $Q\sim 0$, with its sharpness determined by the ``mass term''.  The mass term in the MDC spectrum depends on the ratio $r$, and this generally leads to a smaller  magnitude. Upon turning off the decay rate $\Gamma_0$, 
 both the EDC and MDC spectra reduce to the expected  spectrum $\varepsilon_k = \vv_\Phi \k$, arising from the pole of the auxiliary Greens function in \disp{twin-self}. These expressions  illustrate an  unusual feature of this theory: the two   dispersions  are influenced by the emergent energy scale $\Delta_0$, as well as the width $\Gamma_0$  \disp{width}. 

 The above dispersions exhibit an interesting {\bf kink feature} at $Q=0$ in both spectra.  
 The condition  $Q=0$ locates the kink momentum as
 \beq
 (\k)_{kink} = \frac{\Delta_0}{\vv_\Phi (1-r)}, \label{kinkmomentum}
 \eeq
 it corresponds to occupied momenta provided $r >1$, we will  confine to this case below. For the other case $r< 1$, a kink would arise in the unoccupied side, for this reason we do not pursue it here. 
 For $|Q| \gg\Gamma_0$,  the two dispersions asymptotically become $E^*(k) \sim (\vv_0 + (\vv_0-\vv_\Phi) \, sign({\hat{k}}))  \k$ and $E(k) \sim \frac{1}{2-r} (\vv_\Phi + (\vv_0-\vv_\Phi) \, sign({\hat{k}}))  \k$.  Hence these spectra  exhibit a  change in  velocity (i.e. slope) around $Q\sim 0$ of magnitude $2 (\vv_0-\vv_\Phi) v_F$ for the EDC and the usually larger $  \frac{2}{2-r} (\vv_0-\vv_\Phi) v_F$
  for the MDC spectrum.  The change in slope of the  spectrum  occurs  over a range $\Delta Q \propto\Gamma_0$,  thus  becoming sharper  as  $\Gamma_0$ decreases. 
  
  The value of the EDC energy at the kink is found by substituting $Q=0$ 
 and gives
 \beq
 E^*(k_{kink})= - \frac{r}{r-1} \Delta_0 - \Gamma_0, \label{EDC-kink}.
 \eeq
 
The MDC spectrum shows a kink for $2 \geq r \geq 1$ at the same momentum \disp{kinkmomentum}, with energy
\beq
E(k_{kink}) = 
  - \frac{1}{r-1} \Delta_0 - \Gamma_0 \sqrt{\frac{r}{2-r}}, \label{MDC-kink}
\eeq 
 this  feature   is sharper than in the EDC spectrum since the effective damping is smaller.

When $r >2$, the MDC energy is  real only for $|\k| < (|\k|)_{cutoff}$, where the (negative) momentum
$$(\k)_{cutoff} = (\k)_{kink} + \frac{\Gamma_0}{\vv_ \Phi (r-1)} \sqrt{r(r-2)}.$$ 
For $\k $ beyond the cut off, the root becomes complex implying the loss of a clear peak in the MDC spectrum. Thus the spectrum ``fades'' before reaching the kink momentum  \disp{kinkmomentum}. 
Therefore in this case, the kink is less than ideal, unlike the EDC kink or the MDC kink for $1 \leq r \leq 2$,  which should be visible on both sides of the kink momentum. From \disp{limiting}   we may extract the hole density dependence of all the kink  parameters, while  $\Gamma_0$, determining the kink  width, is given in \disp{width}. 

\begin{figure}[t]
\includegraphics[width=\columnwidth]{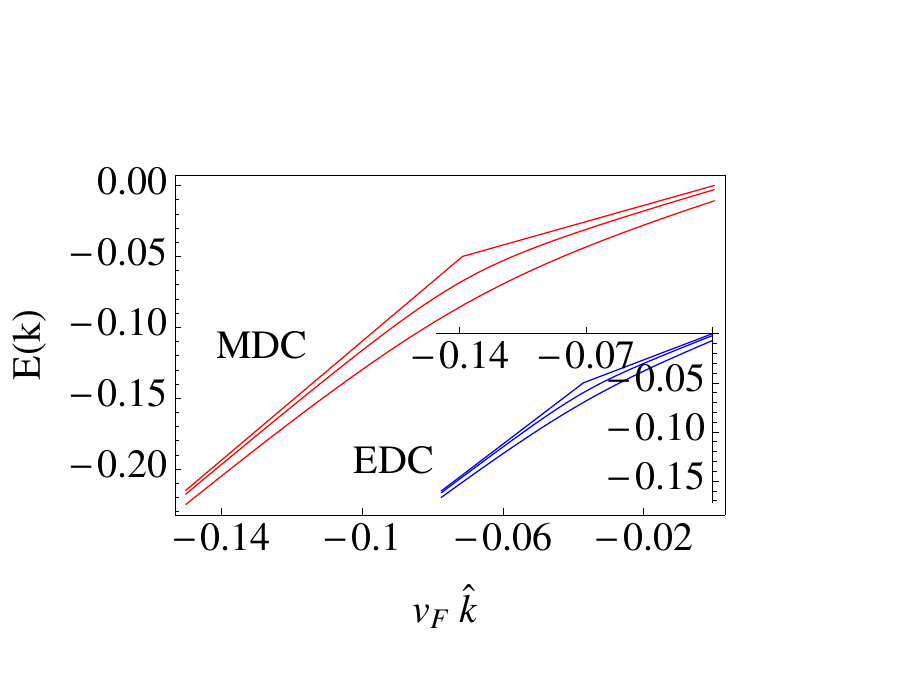}
\caption{ A  kink feature in the  MDC  dispersion relation $E(k)$ from \disp{MDC}  and in the inset from the EDC dispersion $E^*(k)$ \disp{EDC} with parameters $\Delta_0=.025$ eV, $\vv_0=1.05,\vv_\Phi=0.7$ and three values of $\Gamma_0=0.,.01,.02$  in eV from top to bottom. The kink is more pronounced in the MDC curve as discussed in text.} 
\label{fig1}
\end{figure}

We observe in Fig.~(\ref{fig1}) that  the kink becomes sharp when $\Gamma_0$ decreases.   The MDC curves display a sharper kink than the EDC curves, this is easy to understand since the effective damping is smaller in this case, and also the net change in velocity across the kink is greater, as discussed above.
  From \disp{width} we see various parameters that control $\Gamma_0$, in case of laser ARPES, it is argued \cite{Gweon-Shastry} that   $\eta$ is small so we expect to see sharper kinks in this setup. Further, as $T$ drops below $T_c$, the d-wave superconductor has gapless excitations along the nodal direction $<11>$, and the quasiparticles seen in this case are sharper. Theoretical considerations \cite{Shastry-SC} show that in the  superconducting state, a  reduction in the available gapless states responsible for the linewidth   implies a reduction of $\Gamma_0$ and hence to a sharper kink.
 
 We next discuss the feature of  {\bf dynamical asymmetry} in the spectra. It is also important to note that the ECFL spectral function \disp{spectral-function} has an unusual correction to the standard Fermi liquid part, embodied in the caparison function $\mu(k,\omega)$. This function is odd in frequency, thus disturbing the particle hole symmetry of the Fermi liquid part, and it grows in importance as we approach the insulating state since $\Delta_0\to \delta \overline{\Delta}_0$ as indicated in \disp{limiting}. It is also interesting that the spectral line shape in the calculation of Anderson and Casey \refdisp{Anderson-Casey} (AC) as well as Doniach and Sunjic \refdisp{Doniach-Sunjic} (DS) also have such odd in $\omega$ corrections to the Fermi liquid part. In fact the AC result may be viewed as the vanishing of the scale $\Delta_0 \propto k_BT$ so that the ground state is non Fermi liquid like. At finite $T$ and $\omega$ the AC and DS theories are  parallel with the ECFL line shapes regarding the asymmetry as remarked in \refdisp{Asymmetry}, and we wish to make a few comments about the experimental tests for such an asymmetry, going beyond standard measures such as the skewness  factor. 

 DS\cite{Doniach-Sunjic} make the interesting point that the asymmetry is best isolated by looking at the inverse of the spectral function in a plot of  
 \beq
 \frac{1}{A(k,\omega)}  \;\;\; vs \;\;\; (\omega- E^*_k)^2, 
 \eeq
where $E^*_k$ is the peak location in the EDC. With this  plot, a Fermi liquid yields two  coincident straight lines above and below $E^*_k$, whereas an asymmetric contribution, as in \disp{spectral-function} or the DS lineshape \cite{Doniach-Sunjic}, would split into two distinct non linear  curves, from  below and above $E^*_k$.  The inversion of the spectral function is an interesting device, since it refocuses attention on the asymmetric parts. For very similar reasons \refdisp{ECFL} (Fig.~1 inset) also advocates plotting the inverse of the spectral function. On the other hand  an untrained   examination of the EDC curves invariably focuses on the close proximity of the  peaks of $A(k, \omega)$, these are arguably the least interesting part of the  asymmetry story! 
 
In fact armed with the explicit knowledge of the spectral function  of the ECFL theory in \disp{spectral-function}, we can aim to do better in establishing the asymmetry and in determining  the various parameters.
We first redefine   the  frequency by subtracting off  the EDC peak value
\beq
\widetilde{\omega}_k = \omega- E_k^*,
\eeq
so that  the spectral peak occurs at $\widetilde{\omega}_k =0$. The inverse spectral function can be computed as  a function of $\widetilde{\omega}_k$  and reads:
\beq
\frac{A(k,E^*_k)}{A(k,E_k^*+\widetilde{\omega}_k)}=  1+ \frac{e^{u_k}}{2 \Gamma_0} \times \frac{\widetilde{\omega}^2_k}{\Gamma_0 \cosh(u_k) - \widetilde{\omega}_k},\nn \\ \label{ratio}
\eeq
where   the peak value of the spectral function  at $\widetilde{\omega}_k=0$ is :
\beq
A(k,E_k^*) = \frac{A_0}{2} e^{u_k}. \label{peakint}
\eeq
We next construct the object ${\cal Q} (\widetilde{\omega}_k)$ from \disp{ratio} by subtracting unity and cross multiplying:
\beq
{\cal Q}(\widetilde{\omega}_k)& =& \frac{\widetilde{\omega}^2_k}{A(k,E_k^*)/A(k,E_k^*+\widetilde{\omega}_k)-1}.
\eeq
This variable  is  designed to be a $\widetilde{\omega}_k$ independent  constant in a simple Fermi liquid with a Lorentzian line shape  (i.e. \disp{spectral-function} without the caparison function $\mu$).
Here  ${\cal Q}$ has dimensions of the square of energy,  and when plotted against $\widetilde{\omega}_k$  in the small range surrounding zero i.e. $|\widetilde{\omega}_k|\leq \Gamma_0$ it exhibits a linearly decreasing  behavior with $\widetilde{\omega}_k$  within the  ECFL spectral function \disp{spectral-function}
\beq
{\cal Q}(\widetilde{\omega}_k)&=& \Gamma_0^2 (1 + e^{- 2 u_k}) -\left( 2 \Gamma_0 e^{- u_k} \right) \ \widetilde{\omega}_k.
\eeq
 Note that  this function  is flat for the usual Fermi liquid state without asymmetric corrections, since in this case  $u_k \to + \infty$.  If found in data,  this linear in $\widetilde{\omega}$ behavior is the distinctive aspect of the asymmetric lineshapes.  We can then read off various physical quantities once the curve of ${\cal Q}(\widetilde{\omega}_k)$ versus $\widetilde{\omega}_k$ is obtained.  For this purpose we need the intercept ${\cal Q}(0)$ and 
 the slope near  the origin  $\left(d {\cal Q}(\widetilde{\omega}_k)/ d\widetilde{\omega}_k\right)_0 $.
 Clearly the ${\cal Q}(\widetilde{\omega}_k)$ function will deviate from a straight line sufficiently far from $\widetilde{\omega}_k=0$, and it will also be contaminated with background terms as well as noise.  However, with high quality data this procedure could be useful in inverting the data to fit simple functional forms, and to make decisive tests of the predictions of the theories containing asymmetry, namely the DS and AC theories as well as ECFL.

 \section{ Exact formulation in terms of a canonical  Fermions \label{canonical-fermions}} 
 We will next rewrite this in  canonical Fermi representation in {\em an enlarged Hilbert space}  where double occupancy is permitted, and the singly occupied states form a  subspace. We regard  the physical subspace of states $| \Psi\rangle$  as those  that
satisfy the condition of single occupancy, i.e.  $\hat{D}| \Psi\rangle =0$ with the double occupancy operator $\hat{D}$  is given by:
\beq
\hat{D} = \sum_i \hat{D}_i, \;\;  \hat{D}_i \equiv \chd{i \up} \ch{i \up} \chd{i \dn} \ch{i \dn}. \llabel{Ddef}
\eeq
and $\ch{i \si}$ and $\chd{i \si}$ denote the canonical Fermionic destruction and creation operators. The unphysical states contain one or more  doubly occupied states.
 In terms of these,  the Gutzwiller projector over all sites is written as:
\beq
\hat{P}_G= \prod_i \left( 1 - \hat{D}_i \right). \llabel{gutz}
\eeq
This projection operator can be introduced into a partition function to deal with  unphysical states,  as we show below.

The next goal (see Table~\ref{flowchart}) is to write the most efficient representation in the enlarged space  of the \tJ model Green's functions, in terms of the canonical operators and the projection operator.  As pointed out in the Introduction,  we note that pairs of operators that are mutual adjoints in the \tJ model (e.g. $\X{i}{0 \si}= (\X{i}{\si 0})^\dagger$), are allowed to be represented by operators  that  violate this adjoint property. The main result of this section is that  this possibility leads to the most compact canonical theory; we term it the 
non-Hermitean theory. However we first  warmup  with a short summary of  the  more obvious Hermitean theory, which sets the stage for the main result.

\subsection{ A Hermitean canonical  representation with redundancy \label{hermitean}}

 Projected Fermi operators distinguished by the hats can be written in a familiar construction \cite{chats}
\barray
\chl{i \si}&=&\ch{i \si} ( 1- N_{i \sib} ) \nn \\
\chdl{i \si}&=&\chd{i \si} ( 1- N_{i \sib} ), \llabel{cproj}
\earray
where $\n{i \si} = \chd{i \si} \ch{i \si}$, and $\n{i} = \sum_\si \n{i \si}$,
with the property that these conserve the  number of doubly occupied sites locally:
\barray 
~[\chl{i \si} , \hat{D}_i]=0, \;\;
~[\chdl{i \si} , \hat{D}_i]=0. \llabel{localsym}
\earray
 and therefore also globally i.e. with $\hat{D}$ in place of $\hat{D}_i$.  It implies that any Hamiltonian written in terms of these operators with hats commutes with the individual $\hat{D}_i$ as well as the global $\hat{D}$, and thus conserves the local  symmetry of the model. Therefore  acting within  the physical subspace of states, 
 \dispop{cproj} provide a faithful   realization of the $\X{i}{a b}$ operators
as $\X{i}{0 \si} \leftrightarrow  \chl{i \si} $  and  $\X{i}{ \si 0 } \leftrightarrow  \chdl{i \si}$, and clearly satisfies the mutual adjoint property. We are also interested in the product  of two $X$'s in order to represent the kinetic energy term of the effective Hamiltonian below.
  The  optimal choice is seen to be 
\beq
\X{i}{\si 0} \X{j}{0 \si} \leftrightarrow  \chd{i \si} \ch{j \si} \left( 1- N_{i \sib} - N_{j \sib} \right). \llabel{minimal}
\eeq
While  the choice 
\beq \X{i}{\si 0} \X{j}{0 \si} \leftrightarrow  \chdl{i \si} \chl{j \si} \llabel{maximal} \eeq
 is also a faithful representation,  it contains 
an extra term 
$  \chd{i \si} \ch{j \si}  N_{i \sib}  N_{j \sib}$, over and above \dispop{minimal}, 
which is redundant since \dispop{minimal} already commutes with \dispop{gutz}.

 Using \dispop{minimal}  we write a canonical expression for the Hamiltonian 
 \beq H_{tJ} \to \hat{H}_{eff} = \hat{H}_t+ \hat{H}_J, \label{heff}
 \eeq
  with
\beq
&&\hat{H}_t= \hat{T}_{eff} - \chem \sum_i N_{i \si}, \nn \\
&&\hat{T}_{eff} = - \sum_{i j \si} t_{ij}  \chd{i \si} \ch{j \si} \left( 1- N_{i \sib} - N_{j \sib} \right)  \llabel{teff1},
 \eeq
 we call this as  the {\em symmetrized  kinetic energy} in view of its obvious symmetry under the exchange $i \leftrightarrow j$, and  write
$
\hat{H}_J\to \frac{1}{2} \sum_{ij} J_{ij} \left( \vec{S}_i . \vec{S}_j - \frac{1}{4} \n{i} \n{j} \right),
$ 
 with the spin and number operators written in terms of $C$'s and $C^\dagger$'s without hats (since the occupancy of a site is unaffected by the exchange term).
 We easily verify  that
\beq
[\hat{H}_{eff}, \hat{D}]=0=[\hat{H}_{eff}, \hat{P}_G], \llabel{gutzcommute-1}
\eeq
therefore if we start with a state  satisfying $\hat{D} | \Psi \rangle =0 $, i.e.  in the singly occupied subspace,  the resultant  state $H_{eff} |\Psi \rangle $  remains in  this subspace; we  do not create doubly occupied states. 
We note that \dispop{gutzcommute-1}  implies that  the  operator \dispop{gutz} is invariant under time evolution  through $H_{eff}$: 
\beq
\hat{P}_G(\tau) = \hat{P}_G(0). \llabel{gutzcommute}
\eeq
 The partition functional as in \dispop{part1}, now defined  with arbitrary $\tau_0$:
\beq
{Z}=    {\tr}  \  e^{-\beta \hat{H}_{eff}} T_\tau \left( e^{- \A_S} \hat{P}_G(\tau_0) \right), \label{angles}
\eeq
where  the trace (unlike that  in \disp{part1}), is over the entire canonical basis, i.e. includes doubly occupied states. For the observables as well as the source terms $\A_S$, we    use the replacement rules:
\barray
\X{i}{0 \si} \to \chl{i \si}, \;\; \X{i}{ \si 0} \to \chdl{i \si},\;\;\X{i}{\si \si'}\to\chd{i \si} \ch{i \si'}, \llabel{rephats}
\earray
to convert arbitrary expressions involving $\X{i}{ab}$ into those with the $\chl{},\chdl{}$. 
Note  that the density or spin density type variables are replaced by the canonical operators without a hat, since these commutes with the local  $\hat{D}_i$.

We can compute the Green's functions in the enlarged (canonical) basis from 
\beq
\G_{\si_i \si_f}(i \tau_i, f \tau_f) =
    - \frac{\langle T_\tau \left( e^{- \A_S} \chl{i \si_i}(\tau_i) \chdl{f \si_f}(\tau_f) \ \hat{P}_G(\tau_0) \right)\rangle}{ \langle T_\tau \left( e^{- \A_S} \hat{P}_G(\tau_0) \right) \rangle},  \nn \\ \llabel{eq21}
\eeq
evaluated \cite{angleaverage} at ${\A_S \to 0}$. 
This relation  can be replaced by differentiating the partition functional \dispop{angles}  with the Fermi sources $J, J^*$.       Using the commutation of $\hat{P}_G$ or $\hat{D}$ with all operators  and \dispop{gutzcommute}, we are free at this stage to locate  place $\hat{P}_G$  at any specific  time, without affecting the results. This formulation of the theory has parallels with the path integral representation  of the  electromagnetic field (QED) in the temporal gauge, where the scalar potential is chosen to be zero  (i.e. $\phi(r t)=0$)
\refdisp{dirac}, \refdisp{creutz}. In this case the Gauss's law condition $\nabla .\vec{E}(r,t)=0$ needs to be imposed at each time slice. However  upon using $[H,\vec{E}]= \vec{\nabla} \times \vec{B}$,  this object commutes with the Hamiltonian $[H, \vec{\nabla}. \vec{E}]=0$, and therefore it suffices to impose this condition at the initial time. The situation has a clear analogy with \disp{angles}, where it suffices to insert the projection operator at the initial time.

\subsection{ The   Hat Removal Rule  and optimal  Non-Hermitean Theory \label{unhermitean}}

The non-Hermitean theory arises when we inspect closely   expressions of the type in \dispop{angles}, with the time $\tau_0$ chosen as the earliest time $0^-$.  The general argument has been given in the Introduction, we consider its specific application to the present problem next.
Discretizing the  time variables and expanding, we obtain a  series containing expression of the type 
$$\mbox{const} \times  \sum \langle i |  Q_1(\tau_1)\ldots Q_m(\tau_m) \hat{P}_G |i\rangle,$$ 
so that the first  operator from the right $Q_m(\tau_m)$ acts upon a state which is Gutzwiller projected. 
 Now the creation operators contained in the $Q(\tau)$'s are defined with the hats (see \dispop{cproj}) ensuring that they never create doubly occupied states. Next  observe that destroying a particle  cannot { \em create} a doubly occupied site. Therefore it cannot take a projected state out of this  subspace! Therefore the operator $\chl{i \si}$ can as well be replaced by the destruction operator $\ch{i \si}$  {\em without a hat}. We can  iterate this argument for the next operator,  which also acts on a Gutzwiller projected state, and so forth,   leading to the hat removal rules. In this argument,      
we may replace the  operator's $Q(\tau_m)$ by any  expressions involving the destruction operators as well as creation operators with hats (as in \dispop{cproj}), and the same argument holds. More formally we may summarize by saying  that the destruction operator  {\em conditionally commutes}  with the projection operator, when right-operating on projected states:
\beq
[\ch{i \si}, \hat{P}_G] \hat{P}_G=0, \label{quasicommute}
\eeq 
although $[\ch{i \si}, \hat{P}_G] \neq 0$, as one  readily checks. Thus the commutator lives in an orthogonal subspace to that spanned by the Gutzwiller projected states.
 This property also extends to arbitrary functions $\hat{f} $ ($\hat{f} \equiv  \hat{f}\{ \ch{i \si}\}, \{ \chdl{j \si'} \}) $ of the  operators:
\beq
[\hat{f}, \hat{P}_G] \hat{P}_G=0. \label{quasicommute2}
\eeq 
This property is just  a rewriting of  the important block triangularity condition of the operators noted in \disp{eq36} leading to  \disp{eq322}. 
We will  make frequent use of  this expression below.

We now turn to implementing this observation. Let us write
  the partition functional 
\beq
 &&Z= {\tr}  \  e^{-\beta \hat{H}_{eff}} T_\tau \left( e^{- \A_S}  \hat{P}_G(0^-) \right), \label{angles2}
 \eeq
 and  introduce the  important abbreviation for  averages:
 \begin{widetext}
 \beq
 \lll A(\tau_1) B(\tau_2) \ldots \rrr \equiv 
\frac{1}{Z}    {\tr}  \  e^{-\beta \hat{H}_{eff}} T_\tau \left( e^{- \A_S}  A(\tau_1) B(\tau_2) \ldots \hat{P}_G(0^-) \right), \label{gen2}
\eeq
\end{widetext}
where notice that   we located the projector at  the {\em initial time}, by bringing it under the time ordering symbol. 

We now state  the crucial  {\em hat-removal  rule}:  in all expressions of the type \disp{angles2} and \dispop{gen2},
the hats on {\em all  destruction operators} can be removed 
\beq
\chl{i \si}(\tau) \to \ch{i \si}(\tau),  \llabel{unhat}
\eeq
leaving  the  result unchanged.  Notice that this rule can also be applied to  $H_{eff}$,  and the source terms $\A_S$ containing  the destruction operators  $\ch{i \si }$. Note that the {\em creation operators} cannot be `un-hatted' in this fashion- since these do  create a doubly occupied site. 
Summarizing, we can use instead of \dispop{rephats}, the more compact   non-Hermitean rule
 \barray
&&\X{i}{0 \si} \to \ch{i \si}, \; \X{i}{ \si 0} \to \chdl{i \si} = \chd{i \si}(1- N_{i \sib}) , \nn \\
&& \;\X{i}{\si \si'}\to\chd{i \si} \ch{i \si'}.  \llabel{rephats2}
\earray
 We thus     rewrite  the sources  \dispop{sources} as:
\barray
\A_S(i, \tau) &=&  \left(  \chdl{i \si}(\tau)\ J_{i \si}(\tau) + J^*_{i \si}(\tau) \ch{i \si}(\tau)\right) +\nn \\
&& \V_{i}^{\si' \si}(\tau) \chd{i \si'}(\tau) \ch{i \si}(\tau),  \llabel{asource}
\earray
and the   Green's function with imaginary time  $0 \leq \tau_i, \tau_j \leq \beta$
is  therefore written as:
\beq
\G_{\si_i \si_f}(i \tau_i, f \tau_f) =  - \lll \ch{i \si_i}(\tau_i) \chdl{f \si_f}(\tau_f) \rrr,  \llabel{eq23}
\eeq
analogous to \dispop{eq21}  but with   an    unprojected  destruction operator. We will show below that this is the most useful and compact expression for the Green's function. To complete the description of this theory, we turn to the task of  specifying the Hamiltonian, and  obtain the boundary conditions on the time variables. The last task is  somewhat nontrivial since the   projection operator does not commute with the other operators.

\subsection{Hamiltonian in the  Symmetrized and Minimal theories \label{3C} }
In order to represent the Hamiltonian, the spin operators of the  exchange part $H_J$ are unambiguously expressed in terms of the $\ch{i\si}$ and $\chd{i \si}$ operators without hats as in \dispop{rephats2}, since they preserve the occupation of a site. For the kinetic energy
we  could  choose to work with \dispop{teff1},  
and  thereby gain some advantage of dealing with a Hermitean Hamiltonian. This leads to the equations of motion termed the  {\em the symmetrized theory} in \refdisp{Monster}.
Alternately we can implement the hat removal rule for the kinetic energy as well:
\barray 
\hat{T}_{eff} = - \sum_{i j \si} t_{ij}  \chdl{i \si} \ch{j \si}.   \llabel{teff2} \earray
 This minimal version of the kinetic energy is clearly non-Hermitean. However,  it has exactly the 
same action as the symmetrized version  \dispop{heff}, when right-operating on the physical Gutzwiller projected states, as proved above. 
This leads to equations of motion of the {\em minimal theory}  noted in \refdisp{Monster}. 
 and elaborated upon in \refdisp{Large-D} and \refdisp{DMFT-ECFL}. For completeness,  we provide in Sec (\ref{path2}) a brief derivation of these equations for the minimal case, using the above canonical representation, in place of the Schwinger equations.

\subsection{ Kubo-Martin-Schwinger antiperiodic boundary conditions \label{KMS}}
In working with the expression \disp{angles2}, \disp{rephats2} and  \disp{eq23}, we have  assumed that all  the times $\tau_j$ are positive and satisfy $0 \leq \tau_j \leq \beta$.  The Green's function \disp{gdef} satisfies the
Kubo-Martin-Schwinger (KMS) anti-periodic boundary conditions \cite{apbc}
\beq
\G(a\, \tau_i=0,b\, \tau_f)&=&- \G(a\, \tau_i=\beta,b\, \tau_f),\label{27a} \\
\G(a\, \tau_i,b\, \tau_f=0)&=&- \G(a\, \tau_i,b\, \tau_f=\beta), \label{27b}
\eeq
where the fixed time $\tau_f$ ($\tau_i$)  in the first (second) equations is assumed to satisfy   $0 \leq \tau \leq \beta$. These conditions are usually  proven by  using the cyclic invariance of the trace \cite{Kadanoff-Baym}, and translates   easily to the canonical representation  \disp{eq21},  with $\chl{}$ and $\chdl{}$ replacing the $X$ operators \dispop{rephats}.

 In using the non-Hermitean representation \dispop{rephats2}  as in \dispop{eq23}, we cannot use cyclicity of trace since the operator $\chl{}$ does not commute with $P_G$.   Remarkably enough, the   conditional commutativity  \dispop{quasicommute} and \dispop{quasicommute2} suffices to guarantee the required antiperiodicity.
 In physical terms these proofs  follow from the observation  made above, the  creation operators with hats, and destruction operators (without hats)  preserve  a Gutzwiller projected state  within  that subspace.

For simplicity we present the case with sources  turned off i.e. $\AA\to 0$, the more general case follows by a similar argument.  From the definitions of the Green's functions, \disp{27b} is true since $\tr \left(e^{- \beta H_{eff}} \ch{a \si} (\tau_i) [\chdl{b \si'}(0), \hat{P}_G]\right)$ vanishes identically from \disp{localsym}.
 
In order to prove that  \disp{27a}  remains  true, we need to show that  the expression
\beq
\tr \left( e^{- \beta H_{eff}} \chdl{b \si'}(\tau_f) [\ch{a \si} (0), \hat{P}_G] \right) \label{27a1} 
\eeq 
vanishes, despite the non vanishing of the commutator in the expression.  For this purpose, we utilize the conditional commutator \dispop{quasicommute} to write  $[\ch{a \si} (0), \hat{P}_G]=[\ch{a \si} (0), \hat{P}_G] (\iden - \hat{P}_G)$. We next use cyclicity of trace and the simple identity   (for any  $\hat{Q}$):
$\tr\left( (\iden - \hat{P}_G) \hat{Q} \hat{P}_G \right)=0$,  to write the required expression \dispop{27a1} in the form
\beq
\tr \left((\hat{P}_G-\iden) e^{- \beta H_{eff}} \chdl{b \si'}(\tau_f)  \hat{P}_G \ch{a \si} (0) \right). \label{27a2} 
\eeq 
Using $(\hat{P}_G)^2=\hat{P}_G$, we rewrite this as:
$$(\hat{P}_G-\iden) e^{- \beta H_{eff}} \chdl{b \si'}(\tau_f)  \hat{P}_G=[\hat{P}_G, e^{- \beta H_{eff}} \chdl{b \si'}(\tau_f)]  \hat{P}_G.$$  This expression vanishes  on using  the conditional commutator \disp{quasicommute2}, thereby proving the required result \dispop{27a}.

The two canonical theories providing an exact mapping of the original theory are summarized in the Table~(\ref{table2}).

\begin{table*}[t]
\begin{center}
\begin{tabular}{| p {.9 in} |p{ 1.35in} |p{2 in}| p {2.25 in} |}\hline 
& Hubbard-Gutzwiller Theory &(Canonical)  Hermitean Theory & (Canonical) Non-Hermitean Theory \\ \hline
Operators: & $X^{\si 0}$ & $\chdl{\si}= \chd{\si} \ (1- N_{\sib})$ & $\chdl{\si}= \chd{\si} \ (1- N_{\sib})$ \\ \cline{2-4}
& $X^{0 \si }$&$\chl{\si}= \ch{\si} \ (1- N_{\sib})$ & $ \ch{\si}$ \\ \cline{2-4}
& $X_{i}^{\si \si' }$& $\chd{i \si} \ch{i \si'}$& $\chd{i \si} \ch{i \si'}$ \\ \hline
Partition Functional: $Z$ &$\tr_{tJ} e^{-\beta H_{tJ}}  T_\tau \left( e^{- \A_S}  \right)$ &$  {\tr}\, e^{-\beta \hat{H}_{eff}} T_\tau \left( e^{- \A_S} \hat{P}_G(\tau_0) \right)$; \;\;Arbitrary  time $\tau_0$ ($0 \leq\tau_0 \leq\beta$). & ${\tr}\,   e^{-\beta \hat{H}_{eff}} T_\tau \left( e^{- \A_S}  \hat{P}_G(0^-) \right)$\\ \hline
 Green's function: $-\G(1,1')$ &$\langle T_\tau( e^{- \A_s} \X{1}{0 \si} \X{1'}{ \si' 0})\rangle \ $ & $\langle T_\tau \left( e^{- \A_S} \chl{ \si}(1) \chdl{ \si'}(2)  \hat{P}_G(\tau_0) \right)\rangle \ $ \ Arbitrary  time $\tau_0$  ($0 \leq\tau_0 \leq\beta$).&$\langle T_\tau \left( e^{- \A_S} \ch{ \si}(1) \chdl{ \si'}(2)  \hat{P}_G(0^-) \right)\rangle \ $ \\ \hline
Remarks:&$H = H^\dagger$ in  the defining representation. &Symmetrized Theory  $H = H^\dagger$  & (i)Symmetrized Theory: $\hat{H}_{eff} = {H}_{eff}^\dagger$   \\ 
%&&& or \\
&&& (ii) Minimal Theory:  $\hat{H}_{eff} \neq \hat{H}_{eff}^\dagger$ \\ \hline 
\end{tabular}
\caption{ A summary of the the representations of the Green's functions. The non-Hermitean minimal theory provides the most compact set of equations of motion, which are identical to those from the Hubbard-Gutzwiller theory in the second column. The absence of  the adjoint property  for the non-Hermitean theory arises from the asymmetric hat removal  between the  destruction and creation operators in the first two rows of the last column. \label{table2}
}
\end{center}
\end{table*}

\section{The auxiliary Green's function and the caparison function using canonical Fermions  \label{product-ansatz}}
We next discuss the rationale for decomposing the Green's function into an auxiliary Greens function and a caparison function as in \refdisp{ECFL},   using a simple argument from the exact formula \disp{eq23}.  This important part of the theory is also encountered in Section~(\ref{Dyson-Maleev}).
In its  simplest version   this decomposition   can be  illustrated using 
 the  minimal theory, where the averages are defined as
  in \disp{angles2}, with the projection operator pinned at the initial time.   We 
   recall the Green's function from \disp{eq23}  $ \G_{\si_i \si_f}(i \tau_i, f \tau_f) =  - \lll \ch{i \si_i}(\tau_i) \chdl{f \si_f}(\tau_f) \rrr$, with the averages from \disp{gen2}.   Expanding the  $\chdl{}$ operator this becomes
\beq
&&\G_{\si_i \si_f}(i \tau_i, f \tau_f) = - \lll \ch{i \si_i}(\tau_i) \chd{f \si_f}(\tau_f) \rrr +\nn \\
&& \lll \ch{i \si_i}(\tau_i) \chd{f \si_f}(\tau_f) N_{f \sib_f}(\tau_f)\rrr. \label{eq56}
\eeq
We next define  the auxiliary Green's function as:
 \beq
 \GH_{\si_i \si_j}(i \tau_i, j \tau_j) =- \lll \ch{i \si_i}(\tau_i) \chd{j \si}(\tau_j) \rrr ,\eeq
 and regarding the spin, space and time indices as matrix indices with a matrix  inverse $\GHI$.
  By  separating the disconnected and connected parts $(\_c)$ of the second term in \dispop{eq56} we write
  \begin{widetext}
 \beq
\!\!\!\! \lll \ch{i \si_i}(\tau_i) \chd{f \si_f}(\tau_f) N_{f \sib_f}(\tau_f)\rrr=  
 - \GH_{\si_i \si_f}(i \tau_i, f \tau_f)  \langle  N_{f \sib_f}(\tau_f) \rangle  
 +  \lll \ch{i \si_i}(\tau_i) \chd{f \si_f}(\tau_f) N_{f \sib_f}(\tau_f)\rrr_{c}.    \label{discon}
 \eeq
 \end{widetext}
 The connected part  is written in terms of a second self energy $\Psi$ defined as
 \beq
\Psi_{\si_i \si_f}(i \tau_i, f \tau_f)= \GHI_{\si_i \si_{\bf k}}(i \tau_i, {\bf k} \tau_{\bf k})  \times  \nn \\
  \lll \ch{{\bf k} \si_{\bf k}}(\tau_{\bf k}) \chd{f \si_f}(\tau_f) N_{f \sib_f}(\tau_f)\rrr_{c}, \nn \\
\label{Psi} \eeq
 and assembling these we rewrite \dispop{eq56} as the product relation \refdisp{ECFL}
\beq
&&\G_{\si_i \si_f}(i \tau_i, f \tau_f) = \GH_{\si_i \si_{\bf k}}(i \tau_i, {\bf k} \tau_{\bf k}) \mu_{\si_{\bf k} \si_f}({\bf k} \tau_{\bf k}, f \tau_f), \nn \\
&& \mu_{\si_i \si_f}(i \tau_i, f \tau_f) =  \delta(i f)  \left(1- \langle N_{\sib_i}(\tau_i) \rangle\right) +  \Psi_{\si_i \si_f}(i \tau_i, f \tau_f). \nn \\ \label{factor}
\eeq
There is a slight ambiguity in defining the two objects $\GH$ and $\mu$, since we have the freedom of adding a common function to the two parts of \disp{eq56} that cancels out in the physical Greens function. Apart from this,  we expect that the two  objects in \disp{factor}   are exactly equivalent to  the auxiliary Greens function and the caparison factor in \disp{factor2}, \disp{caparison} and \disp{auxg}  as found from the Schwinger method.

We observe from the expression \dispop{Psi}  that  if  the averages  are (temporarily) computed   in a standard Feynman Dyson theory,  then  $\Psi$ is essentially the  self energy of a Hubbard type model,  made dimensionless by dropping  an explicit interaction constant $U$. Indeed this is the key observation made  in \refdisp{ECFL},  on the basis of the $\lambda$ expansion, where  the two self energies are argued to be generically Fermi liquid-like and  similar to each other.  An   energy scale ($\Delta$) emerges from a ratio of their imaginary parts, and    controls the significant asymmetry seen in  the spectral functions.

\section{The $\lambda$-Fermions \label{lambda-Fermions}}

A  natural question is whether   \disp{Min-2}, explicitly containing the parameter $\lambda$, can arise  in  a microscopic  theory  where $\lambda$ enters in a fundamental way, as opposed to the ``engineering approach''  in Section~(\ref{second-chem}). 
 A set of  $\lambda$-Fermi operators are defined below,  as generalized version of the non-Hermitean representation \dispop{rephats2}  with a  parameter $\lambda \in [0,1]$ providing a continuous interpolation between the free Fermi and extremely correlated limits:
\barray
&&\X{i}{ \si 0 }(\lambda) \to \chd{i \si} ( 1- \lambda \chd{i \sib}\ch{i \sib} ) \nn \\
&&\X{i}{ 0 \si  }(\lambda) \to \ch{i \si}  \nn \\
&&\X{i}{ \si \si' }(\lambda) \to \chd{i \si} \ch{i \si'} \ . \label{lambdaxs}
 \earray
  Clearly  $\lambda=0$ gives us back the canonical Fermion operators, whereas $\lambda=1$ gives the Gutzwiller projected Hubbard $X$ operators \refdisp{Hubbard} as in \dispop{rephats2}, provided the states are Gutzwiller projected. A feature of this representation  is the loss of the adjoint property, i.e. $\left(\X{i}{ \si 0 }(\lambda)  \right)^\dagger \neq \X{i}{ 0 \si  }(\lambda) $, unless $\lambda =0$.

 These operators satisfy a  $\lambda$  dependent (graded)  Lie algebra with fundamental brackets that are partly Fermionic and partly Bosonic. Using the
 canonical anticommutation relations of the $\ch{},\chd{}$ operators, 
  we work out the fundamental Fermionic bracket:
 \beq
 \{ \X{i}{0 \si_i}(\lambda), \X{j}{\si_j 0}(\lambda) \} &=& \delta_{ij} \{ \delta_{\si_i \si_j} - \lambda \,  \si_i \si_j \X{i}{ \sib_i \sib_j }(\lambda)  \}. \nn \\ \label{lambda-reln2}
 \eeq
  We next evaluate { the fundamental Bosonic bracket}
 \beq
~[  \X{i}{0 \si_i}(\lambda), \X{j}{\si_j \si_k}(\lambda)]&=&\delta_{i j} \delta_{\si_i \si_j} \X{i}{0 \si_k}(\lambda) \label{lambda-reln3}
 \\
~[  \X{i}{ \si_i 0 }(\lambda), \X{j}{\si_j \si_k}(\lambda)]&=&- \delta_{i j} \delta_{\si_i \si_k} \X{i}{ \si_j 0}(\lambda)  \label{lambda-reln4}.
 \eeq
 Here \dispop{lambda-reln4}   requires a brief calculation \cite{calc-commute}   invoking the Pauli principle vanishing of  $\chd{\si}\chd{\si}\to0$.  On the other hand  \dispop{lambda-reln3} is elementary, due to the absence of $\lambda$ in both sides of the equation.   At $\lambda=1$ these reduce to  the relevant  subset of  the Hubbard algebra  \refdisp{Hubbard} found from the fundamental definition $\X{i}{ab}= |a\rangle \langle b| $ .

  The representation \dispop{lambdaxs}  does not  at general $\lambda$ reproduce the ``half bracket'', or product  relations expected for {\em projection operators}. We find that
  \beq 
  \X{i}{ \si 0 }(\lambda)\X{i}{ 0 \si'  }(\lambda) \neq \X{i}{ \si \si' }(\lambda), 
  \label{not-proj} \\
\X{i}{ 0 \si'  }(\lambda)  \X{i}{ \si 0 }(\lambda) \neq \X{i}{00} \delta_{\si \si'}.  \label{not-proj2}
   \eeq
The   exceptions are at $\lambda=0$, where it is trivially true, and non trivially 
   at  $\lambda=1$, where Gutzwiller projection of the allowed states does restore this property when right-operating on the projected states. In the Green's functions below, we will equate the {\em averages} of both sides of  \disp{not-proj}. This equality of the averages acts  as the number  constraint and fixes the chemical potential $\chem$. In doing so, the average of \disp{not-proj2} is not constrained and takes on a suitable value  determined by the   anticommutation relation \disp{lambda-reln2}.
      
 This representation can be used to define  a many-body problem where the $\lambda$ dependent EOMs for the  Green's functions constructed from \dispop{lambdaxs} can be written down. Observe that  the EOMs for the Green's functions only require the use of  \dispop{lambda-reln2} and the Heisenberg equations of motion,  and in turn these arise from the basic Lie commutators (anticommutators) of the type given in \dispop{lambda-reln3} and \dispop{lambda-reln4}. The calculation does not   ever require the use of  product relations of the type \dispop{not-proj}.
 It then  follows that we  can replace the \tJ Hamiltonian  and the operators in the original theory by their $\lambda$-versions, i.e. replacing $\X{i}{ab} \to \X{i}{ab}(\lambda) $, and thereby obtain equations that yield \dispop{Min-2}.  This procedure then provides a (continuous) interpolation between the free Fermi and extremely correlated regimes by varying $\lambda$ from $0$ to $1$. Let us first demonstrate this by a brief calculation.
 
 \subsection{ The $\lambda$-Fermion theory equations of motion. \label{secC}}
 Using the $\lambda$ Fermions, we define the Green's function as
\beq
\G^{(\lambda)}_{\si_i \si_f}(i \tau_i, f \tau_f) = - < T_\tau \X{i}{0  \si_i}(\tau_i, \lambda ) \X{f}{\si_f 0} (\tau_f, \lambda )  >_{(\lambda)} \; \label{eq460}
\eeq
where with arbitrary $\hat{A}$
\beq
<\hat{A}>_{\lambda}&\equiv&    -  \frac {\tr \ e^{-\beta H_{eff}(\lambda)} T_\tau \left( e^{- \A_S(\lambda)} \hat{A} )  \right)}{Z(\lambda)}, \nn \\
Z(\lambda)&=& {\tr \ e^{-\beta H_{eff}(\lambda)} T_\tau \left( e^{- \A_S(\lambda)} \right)}.\!\!\!\!  \llabel{eq47}
\eeq
In this expression  $H_{eff}(\lambda)$ is given by \disp{hamiltonian-lambda} and  $\A_S(\lambda)$  is obtained from  \dispop{sources}, with the replacement   $\X{i}{ab}\to \X{i}{ab}({\lambda})$:
\begin{widetext}
\barray
H_{eff}(\lambda)&=& - \sum_{i j } t_{ij} \X{i}{\si 0}(\lambda)\X{j}{0\si}(\lambda) - \chem \sum_i N_{i \si } + \lambda \frac{1}{2} \sum_{ij} J_{ij} \left( \vec{S}_i . \vec{S}_j - \frac{1}{4} N_{i \si} N_{j  \si'}   \right)+   u_0 \ \lambda \ \sum_i N_{ i \up} N_{i \dn} . \label{hamiltonian-lambda}
\earray 
 \end{widetext}
  where $u_0$ is now  the ``second chemical potential''.  The scaling of the $J$ term with $\lambda$ is optional, and done here so that we obtain the Fermi gas at $\lambda=0$. Using \disp{lambdaxs}, we see that this  Hamiltonian is linear in $\lambda$ and interpolates between the free Fermi gas and the fully interacting model, when acting on suitably projected states.
The equation of motion of $\G^{(\lambda)}$ can be obtained using  the commutation relations  Eqs~(\ref{lambda-reln2},\ref{lambda-reln3},\ref{lambda-reln4}),  
the calculation  is parallel  to that in  Appendix~(\ref{Minimal}).  In brief, \disp{a} and \disp{aa} are unchanged by working with $\X{}{}(\lambda)$'s, and  in place of \disp{EOM-421} we obtain 
\beq
&&\GHI_{0, \si_i,  \si_j}(i \tau_i,  j \tau_j) \G^{(\lambda)}_{\si_j \si_f}(j \tau_j , f \tau_f)= \nn \\
&&  \delta(\tau_i-\tau_f) \delta_{ij} (1- \lambda \ \gamma_{\si_i \si_f}(i \tau_i) ) \nn \\
&& - \lambda \ \sum_{j \si_j} t_{ij} (\si_i \si_j) \   \langle T_\tau \left(\X{i}{\sib_i \sib_j}(\tau_i)  \X{j}{0 \si_j}(\tau_i)  \ \X{f}{\si_f 0}(\tau_f) \right)\rangle_{(\lambda)} \   \nn \\
&&+ \frac{1}{2} \sum_{j \si_j} J_{ij}  (\si_i \si_j) \langle T_\tau \left(   \X{j}{\sib_i \sib_j}(\tau_i)   \X{i}{0 \si_j}(\tau_i)  \X{f}{\si_f 0}(\tau_f) \right)\rangle_{(\lambda)} \nn \\ 
&&- \frac{1}{2} \lambda u_0  \sum_{ \si_j}   (\si_i \si_j) \langle T_\tau \left(   \X{i}{\sib_i \sib_j}(\tau_i)   \X{i}{0 \si_j}(\tau_i)  \X{f}{\si_f 0}(\tau_f) \right)\rangle_{(\lambda)} \ 
,\nn \\ \label{EOM-4211} 
\eeq
where the $\lambda$ dependence of the $\X{}{}$ operators is implicit.
The higher order Green's functions  may be expressed as functional derivatives with respect to the Bosonic source $\V$,  in the same fashion as in the Appendix~(\ref{Minimal}).
The exchange term $J_{ij}$ does not pick up a factor of $\lambda$ through the EOM 
since it conserves double occupancy. We can choose to additionally scale it with $\lambda$ as $J_{ij} \to \lambda J_{ij} $, so that at $\lambda=0$ we obtain the Fermi gas. This choice  seems reasonable in the liquid phase of the electrons,  in other  phases it is easy enough to recover from this scaling if needed.  To save writing the $u_0$ term  is  absorbed  as $J_{ij} \to J_{ij} - u_0 \delta_{ij}$, with this the  resulting equation is
\beq
  \left(  \GHI_{0, \si_i,  \si_j}(i \tau_i,  j \tau_j) - \lambda \ \hat{X}_{ \si_i \si_j}(i \tau_i, j \tau_j)- \lambda \  {Y_1}_{ \si_i \si_j}(i \tau_i, j \tau_j)\right) \nn \\
  \times \G^{(\lambda)}_{\si_j \si_f}( j \tau_j, f \tau_f) = \delta_{if} \delta(\tau_i-\tau_f)  \left( \delta_{\si_i \si_f} - \lambda \ \gamma_{\si_i \si_f }(i \tau_i) \right).  \nn \\ \label{EOM-422}
 \eeq
The constitutive relation determining the chemical potential is taken as
\beq
n_{i \si} & =& < \X{i}{ \si 0 }(\tau, \lambda) \X{i}{0 \si}(\tau^-, \lambda)>_{(\lambda)}, \nn \\
 &=&  \G^{(\lambda)}_{\si \si}( i, \ \tau^-, \tau), \label{eq450}
\eeq
rather than $n_{i \si}  = < \X{i}{ \si 0 }(\tau, \lambda) \X{i}{0 \si}(\tau, \lambda)>_{(\lambda)}$ (\refdisp{wrong-count}). This  limiting process   corresponds to enforcing the half bracket relation  \disp{not-proj} {\em as an average}.    \disp{eq450} is exact   for the fully projected operators where  $\lambda=1$, while  for other values of $\lambda$  it is guided by the requirement of continuity in $\lambda$.  In the same spirit, we express the function $\gamma$ in \disp{Min-2} as
\beq
\gamma_{\si \si'}(i \tau) &=& \si \si' \G_{\sib' \sib}(i \tau^-, i \tau), \label{gamma-prob2}
\eeq
while the direct computation using \disp{lambda-reln2} would yield identical times, rather than the split times  in \disp{gamma-prob2}. An iteration scheme for solving  these equations using ideas of the skeleton expansion is detailed in  \refdisp{Monster} and in \refdisp{Hansen-Shastry}, and hence we skip the details.

A very simple example can be given to  illustrate the role of $\lambda$ and $u_0 $,  where the skeleton expansion is avoided. Let us  consider the atomic limit of the   $\lambda$-Fermions theory. 
We consider the Hamiltonian $H_0= - \chem \sum_\si N_\si+ \lambda u_0 N_{\up} N_{\dn}$ with $u_0 \geq 0$. The Green's function in \disp{eq460} can be calculated easily using the EOM technique  as: 
\beq
\G(i \omega_n) = \frac{1-n_{\sib} }{i \omega_n + \chem } +\frac{(1-\lambda)n_{\sib}}{i \omega_n + \chem - \lambda u_0}.
\eeq
At $\lambda=0$ or $1$,  this yields the exact atomic limit result, and provides a smooth interpolation between these limits. The positive energy pole at $\lambda u_0 - \chem$ does not contribute to the occupancy for a sufficiently large  $u_0$ and low $T$. In the  more realistic case with non zero hopping discussed in \refdisp{Monster} and \refdisp{Hansen-Shastry}, the energy $u_0$ is non-trivially
fixed by a second sum rule \disp{second-sumrule}, and the iteration procedure is more complex, involving the skeleton expansion.  While  the atomic limit example is quite explicit, it does not generalize in any simple way to the case of finite hopping, and therefore is somewhat trivial.

We next remark on some   consequences of the $\lambda$ expansion in the intermediate region  $\lambda <1$, that follow from general principles. Let us first summarize  the high frequency limit of the Green's functions.
 When $ i \omega_n \to \infty$, the local  Green's function  falls off  as  $\G( i\omega_n)\to a_G/i \omega_n$. Here  the constant $a_G= \langle \{ \hat{C}, \hat{C}^\dagger\} \rangle$, with $\hat{C},\hat{C}^\dagger$  the two appropriate operators involved in $\G$, it  is  a measure of the total fraction of states. In the Hubbard model $a_G=1$, since we have canonical operators, and implicitly $|\omega_n | \gg U$ as well. However  for the \tJ model we obtain $a_G = (1- n/2)$, with a net deficit of $n/2$ states  from the Hubbard model. This deficit is accounted for by  the   upper Hubbard band  that is ignored in the \tJ model. The  lower Hubbard band thus contains a fraction $1-n/2$ of all the  states, of which we account for $n/2$ as the occupied states (with two  spin projections available), and $1-n$ as the unoccupied part of the lower Hubbard band. These $1-n$ states are available for charge excitations in the \tJ model, and freeze out towards the insulating limit.  Summarizing, in this  picture  we  have $n/2$ occupied and $1-n$ unoccupied states in the lower Hubbard band, and $n/2$ states at high energy of $O(U)$.

In the   $\lambda$ expansion, from \disp{Min-2}  we have $a_G= 1- \lambda \gamma$, where $\gamma$ is further  expanded in $\lambda$ .  On enforcing the number sum rule \dispop{eq450} we find that the effective  number of states described by this theory can be decomposed  into 
$n/2$ occupied states and $(1-n)+ (n/2- \lambda \gamma) $ unoccupied states. These are to be taken as    the low energy sector of a fiduciary Hamiltonian. The fraction $ (n/2- \lambda \gamma) $  vanishes only when $\lambda=1$ and is otherwise 
an unspecified surplus of states in the low energy sector.  An  unbalanced state count  of this type is to be  expected when we have non-unitary evolution. Indeed in the second order $\lambda$ expansion carried out numerically,  a similar  excess of states is found \cite[Section ~(2), last paragraph]{Hansen-Shastry}.
 Another related consequence  is that the spectral function positivity, requiring unitary evolution,  can no longer  be  guaranteed- in finite orders of the $\lambda$ expansion.  This feature is  well  recognized in \refdisp{ECFL}, where it is noted that the occupied states with $\omega <0$  are essentially unaffected by this problem.

\section{Analogy with the Dyson-Maleev representation of spin operators \label{Dyson-Maleev}}

\begin{table*}[t]
\begin{center}
\begin{tabular}{| p{1.3 in}|  p{.75in}  | p{1.9 in} |  p {.5in} | p{1.93 in} |}\hline 
%\begin{tabular}{| p{1.5in}|  p{1.5in}|  p{1.95in}|  p{1.5in}|   p{1.95in} |}
%\multicolumn{5}{c}{\bf Table 1.   }  \\ \hline
&\multicolumn{2}{c|}{\bf Spins: The Dyson-Maleev mapping}   &\multicolumn{2}{c|}{\bf Fermions: The non-Hermitean  mapping}  \\ \cline{2-5}
Destruction operator&$S_i^-$&$b_i$&$X_i^{ 0\si }$& $C_{i \si}$\\ \cline{1-5}
Creation operator&$S_i^+$&$(2 s) ~ b^\dagger_i (1-\frac{n_i}{2 s})$&$X_i^{ \si 0}$& $C^\dagger_{i \si}(1- \lambda N_{i \sib})$\\ \cline{1-5}
Density operator(s)&$S_i^z+s$&$  n_i= b^\dagger_ib_i $&$X_i^{ \si \si'}$& $C^\dagger_{i \si} C_{i \si'}$\\ \cline{1-5}
Projection Operator&$\hat{P}_{D}$&$  \prod_{i} \{ \sum_{m=0}^{2s}\delta_{n_i,m}\} $&$\hat{P}_G$& $\;\;\; \prod_i (1-\! N_{i \up} N_{i \dn})$,   for  $\lambda=1$\\ \cline{1-5}
Vacuum& $| \dn \dn \ldots \dn \rangle$ & $|00\ldots0\rangle$&$|Vac\rangle$ & $|00\ldots0\rangle$ \\ \cline{1-5}
Small Parameter \& Its Range&$\frac{1}{2s}$& $ \frac{1}{2s}\in [0,1]$&  $\lambda$  & $\lambda\in [0,1]$\\ \hline
Auxiliary Green's function &&$\GH(i,j)\!=\!-\lll b_i b^\dagger_j\rrr$&& $\GH(i,j) =- \lll \ch{i \si} \chd{j \si}\rrr$ \\ \hline
Caparison Function& &$\mu(i,j)\!=\!\delta_{ij}(1-\frac{1}{2s}\langle n_j\rangle)+\!\frac{1}{2s} \Psi(i,j)$&&$\mu(i,j) = \delta_{ij} (1 -  \lambda \gamma )+ \lambda \Psi(i,j)$ \\ \hline
Second Self energy $\Psi$ &&$\Psi(i,j)=\GH^{-1}(i, {\bf a}) \lll b_{\bf a} b_j^\dagger   n_j \rrr_c$&&$\Psi(i,j)=\GH^{-1}(i, {\bf a}) \lll \ch{{\bf a } \si}  \chd{j \si}   N_{j \sib} \rrr_c$ \\ \hline
\end{tabular}
\caption{A comparison of the Dyson-Maleev representation for spins and the 
non-Hermitean representation \disp{rephats2} for two component Fermions $\si = \pm 1$  with $\sib = -\si$. At $\lambda=1$ the Fermion mappings  provide a faithful representation of Gutzwiller projected Fermi operators $X_i^{ab}$,  {\em acting to the right}  on states with single occupancy, since  their action    produces states  that  remain in this space. The representation is non self adjoint, i.e. its left operation on Dirac bra states is not faithful. The situation has an exact parallel in the Dyson Maleev representation. The Dyson projection operator $\hat{P}_D$ for integer $2 s$  and the Gutzwiller projection operator  $\hat{P}_G$  at $\lambda=1$, play a similar role in filtering out unphysical states. 
 The role of the parameter $\lambda$ away from $0,1$ is similar to that of $\frac{1}{2s}$,    extending the Dyson Maleev representation to spin values that are neither integer or half integer.   The last three rows show the auxiliary Green's function, the caparison function and the second self energy in terms of the Bosons from Eqs~(\ref{eq566}, \ref{eq57}). These follow from the work of Harris, Kumar, Halperin and Hohenberg \refdisp{HKHH} adapted to the ferromagnet.
  The   corresponding Fermionic objects are discussed in Section~(\ref{product-ansatz}) and detailed in Eqs~(\ref{Psi},\ref{factor}). \label{table3}
  }
\end{center}
\end{table*}

The non-Hermitean representation in \disp{rephats2} of the Gutzwiller projected electron operators,  when used with the averaging in \disp{gen2},  was shown in Section~(\ref{unhermitean}) to provide an exact mapping of the \tJ model.   Reflecting on this result,    the author  realized   recently   that  the mapping  \disp{rephats2} is  the Fermionic analog of the  Dyson-Maleev representation for spin operators  \cite{Dyson,Maleev}, used to understand spin wave interactions in magnets (see Table (\ref{table3})).

With the advantage of   hindsight, this connection  seems natural. The Gutzwiller projected electronic  $\X{}{ab}$  operators defined by Hubbard \refdisp{Hubbard},  generate a non canonical algebra  of Fermions that is (partly) given in  Eqs~(\ref{lambda-reln2},\ref{lambda-reln3},\ref{lambda-reln4}) with $\lambda=1$. On the other hand
 the  spin operators   provide the   best studied non canonical Bosonic algebras.   The spins are not quite Bosons,  they are equivalent to ``hard core''  Bosons- with infinite on site repulsion, in parallel to the infinite $U$  in the extremely correlated electron problem.  
 In order to avoid dealing with the infinite energy of the hard core, several other representations of spins were invented, such as the Holstein Primakoff method \refdisp{HP}. Dyson's   use of   a non-Hermitean representation provides the most compact canonical description of the spin operators. In fact it is  analogous to  the non-Hermitean mapping of the Fermionic Gutzwiller problem in \disp{rephats2}.

Dyson's representation,  later streamlined   by Maleev \cite{Maleev},   may be written with $n_i = b_i^\dagger b_i$ as
\beq
&&S_i^+= (2 s)\  b_i^\dagger \ (1- \frac{n_i}{2s}  \ )  \nn \\
&&S_i^- = b_i \nn \\
&&S_i^z +s= n_i , \llabel{dyson}
\eeq
where $\vec{S}_i.\vec{S}_i= s(s+1)$ and $b_i$, $b_i^\dagger$ are canonical Bose operators. 
The Boson vacuum state $b_i |vac\rangle=0$  is mapped as $|vac\rangle \leftrightarrow |\dn,\dn,\dn\ldots \dn\rangle$, so that the action of $b_i^\dagger$ creates spin reversals. Their   number   is   cut off such that $n_i \leq (2s)$, thereby defining the physical states.
Under these conditions \disp{dyson} is shown to provide a faithful representation of the angular momentum operators, when right-operating on physical states.  Under the action of the operators in \dispop{dyson}, the physical states  form an invariant  subspace of the extended Bose Hilbert space, and are selected by  projection. The Dyson projection operator  $\hat{P}_D$ acts on the Bose state space and leaves the physical states unchanged while annihilating states with $n_i >  (2s)$. 
  
 It is now evident that  the Dyson-Maleev representation
has a strong  formal similarity to   the minimal representation \dispop{rephats2}.
The Dyson projector $\hat{P}_D$  plays a role  parallel to that of  the Gutzwiller projector  $\hat{P}_G$ in \dispop{rephats2} in our theory.  The parallel further deepens in the path integral representation of the Fermions that we discuss below.  The  interesting work of Douglass \cite{Douglass},  following Langer's \cite{Langer} path integral program for Bosons- employs the projection operator $\hat{P}_{D}$ in the same spirit to our usage below.

The   work of Harris, Kumar, Halperin and Hohenberg (HKHH) \refdisp{HKHH} extended Dyson's method  to  two sublattice antiferromagnets, and  provided  a  non trivial generalization  to  study   the lifetime of the excitations.  Details of the ECFL formalism   turn out to have  points of   overlap with those in   HKHH that are worth noting.  In particular HKHH decompose the physical Green's function into a space time   convolution of two parts. These parts   are precisely  the Bosonic analogs  of the ECFL breakup of   the physical Green's function,  into an auxiliary Green's function $\GH(k)$ and a caparison function $\mu(k)$, as   detailed in \refdisp{ECFL} and in Section~(\ref{product-ansatz}).

The   computation of  the Green's function by  HKHH \cite{HKHH} was carried out for the two sublattice antiferromagnet. In order to avoid dealing with the added complexity of the  two sublattice problem, we translate their method to the Dyson problem of the dynamical Green's function of the ferromagnet. We  use a    notation that brings out the close parallel  with the product {\em ansatz} used in ECFL \refdisp{ECFL}. 

The  calculation, paraphrasing that  of  HKHH,  proceeds as follows.
In order to compute the imaginary time  Green's function $\G(i,j) =- \lll  S_i^- S_j^+ \rrr $ with the shorthand spacetime notation $i\equiv (r_i, \tau_i)$, the repeated index summation (integration) convention and denoting the averages as  $\lll Q\rrr = \tr (e^{- \beta H} T_\tau Q P_D )/ \tr (e^{-\beta H }P_D)$, we write from \dispop{dyson}
\beq
\frac{1}{2s}\G(i,j) & = & - \lll b_i b_j^\dagger (1 - \frac{1}{2s} n_j \rrr  \label{gspin}
\eeq
Separating out the disconnected part we write  $\lll b_i b_j^\dagger n_j\rrr = \lll b_i b_j^\dagger\rrr \langle n_j\rangle + \lll b_i b_j^\dagger n_j\rrr_c  $, and defining the auxiliary Green's function $\GH(i,j) =  - \lll b_i b_j^\dagger \rrr$ as well as its inverse through $\GH(i,{\bf k}) \GH^{-1}({\bf k},j)= \delta(i,j)$, we arrive at 
 \beq
\frac{1}{2s}\G(i,j) & = & \GH(i,j) (1-  \frac{1}{2s} \langle n_j \rangle) + \frac{1}{2s} \GH(i,{\bf k})\ \Psi({\bf k},j), \;\;\;\;\;\; \label{eq566} \\
\Psi(i,j)  &=&  \GH^{-1}(i, {\bf a}) \lll b_{\bf a} b_j^\dagger   n_j \rrr_c \;\; .  \label{eq57}
\eeq
We use a notation with sums over repeated bold indices everywhere.
 We can rewrite \dispop{eq566} as a convolution of the auxiliary Green's function $\GH$ and a caparison function $\mu$,  in the form   $\frac{1}{2s}\G(i,j)= \GH(i,{\bf k}) \mu({\bf k},j)$, where $\mu(i,j) = \delta_{ij} (1 -  \frac{1}{2s} \langle n_j \rangle )+ \frac{1}{2s} \Psi(i,j)$.  The auxiliary Green's function is defined in terms of its own self energy $\Phi$ through the usual Dyson equation  $\GH^{-1}(i,j)= \GH_0^{-1}(i,j) - \Phi(i,j)$.  Thus the physical Green's function $\G$ is determined in terms of the two self energies $\Phi (k,\omega)$ and $\Psi(k,\omega)$. 
Written in $(k,  \ i\omega)$ space, this is identical to the functional  form in ECFL theory \disp{twin-self}!

 The   corresponding Fermionic objects are discussed in Section~(\ref{product-ansatz}) and detailed in Eqs~(\ref{Psi},\ref{factor}). On comparing the two we recognize that the structure of  Eqs~(\ref{eq566},\ref{eq57}) is the exact parallel of the ECFL theory for the Green's function written  in the  notation of \refdisp{ECFL}.  In the HKHH paper,  the   objects evaluated  amount to  these two   ECFL self energies,  by the correspondence   $\Psi(k,\omega) \leftrightarrow \Lambda(k,\omega)$ (see \cite[Eq~(C10)]{HKHH}), and     $\Phi(k,\omega)\leftrightarrow\Sigma(k,\omega)$ (see \cite[Eq~(2.22)]{HKHH}).  It is worth noting further   that the role of the parameter  $\lambda$ in the ECFL theory is in close parallel to that of $\frac{1}{2s}$ in the magnon problem.  Expansions in these two ``small parameters''  serve to organize the calculations.

   The product ansatz in ECFL \cite{ECFL, Monster} was originally arrived  at in \refdisp{ECFL}  by analyzing the Schwinger equations and insisting on a canonical Green's function to be factored out  from the physical $\G$.  The calculation of HKHH, on the other hand,  was  through a different route  using insights from   the  Feynman diagrams    applied to the four Boson operators in \dispop{gspin}.  It  is satisfying    that the two independent calculations,  one for Gutzwiller   projected  Fermions and the other for  hard-core Bosons,  lead to such  a close parallel, expressed most naturally  in the twin self energy representation \disp{eq566} and \disp{eq57}.   
   
     A few additional comments on the role of the projection operator in the two problems   are relevant here.  Dyson demonstrated  in  his non-Hermitean representation  that magnon interactions at low temperatures lead to   $T^4$ type  corrections to  the magnetization of the  ideal spin wave theory. He argued that the projection operator $\hat{P}_D$ is largely irrelevant in the ferromagnet, and provided an estimate of corrections to the low T behavior arising from this neglect. For the antiferromagnet, HKHH similarly argued  that the projector is unimportant at low $T$, and gave an  estimate of the expected corrections. The corrections are  larger than in the ferromagnet,  and yet smaller than most  quantities of interest at low $T$.   The density of excitations is small at low $T$ in the magnetic problem, and thus  provides a basis  for ignoring the projection operator. However  
 in the Fermion problem studied here, the particle density is never too small in the interesting regime, 
  and hence the projection operator  must be respected. 
Interestingly enough, the projector does not explicitly appear in the Schwinger EOM \disp{Min-2}, but it does determine  the choice of the correct constitutive relation \disp{number-sumrule}.  Thus  the projection operator plays a significant role in enforcing the Luttinger-Ward theorem \cite{lw} for the volume of the  Fermi surface.

 Another major  difference between the Fermionic and the spin problems is the role of the second Lagrange multiplier $u_0$, when the parameter $\lambda <1$. In the Fermi problem, it is essential to change the Hamiltonian by adding the term $\lambda u_o \sum_i N_{i \up} N_{i \dn}$,  in addition to replacing the projected $\X{i}{ab}$ by $\X{i}{ab}(\lambda)$. This is required  in order to satisfy the shift identities, and as explained in \refdisp{Monster}, the parameter $u_0$ is fixed by a number sum rule on the auxiliary Green's function. The problem of  magnetic excitations  does not have a counterpart to this term. However,  we can imagine extending  the  Dyson-Maleev and HKHH formalism to an extremely correlated Bose liquid with a fixed number of Bosons, e.g. $^{4}He$ on a suitable substrate giving rise to a lattice model with hard core repulsion. In such a case, a corresponding  theory parallel to ECFL can be developed, requiring both the shift identities and a second Lagrange multiplier $u_0$  disfavoring multiple occupancy to satisfy these.

 \section{  Path integrals. \label{pathintegral}}
\subsection{ Canonical Path integral representation \label{path1}}
We now introduce path integrals to represent the partition functional \disp{angles2}, wherein the operators are replaced by anticommuting c-numbers, i.e. the Grassman variables. We will keep the discussions to a minimum since  excellent references can be consulted for details \cite{Zinnjustin,Shankar,Popov,Orland}.
We map the operators as
$\ch{i \si} \to c_{i \si}$, $\chl{i \si} \to \widetilde{c}_{i \si}\equiv c_{i \si} ( 1- c^*_{i \sib}c_{i \sib})$,
$\chd{i \si} \to c^*_{i \si}$, $\chdl{i \si} \to \widetilde{c}^*_{i \si}\equiv c^*_{i \si} ( 1- c^*_{i \sib}c_{i \sib})$. The time dependence is dealt with using a standard Trotter decomposition of the non commuting pieces \cite{Orland}.
Handling the Gutzwiller projector is  discussed below and in  
  Appendix \ref{pathint}.  It is understood that when the Trotter index  $M$ is finite, we have a discretized time representation, so that  when $M \to \infty$, we obtain   the continuous time path integrals.  We  work initially with the discrete time version since somewhat subtle identities such as the Pauli principle and the Gutzwiller projection  identities  can be verified explicitly. 
 We now  write the partition functional Z   \dispop{angles2}, in terms of Grassman variables at discrete times $c_{ i \si}(\tau_j)$ and $c^*_{ i \si}(\tau_j)$, and a global integration over all  Grassman variables with the conventional definition \cite{Orland}:
 \beq
&&Z^{(M)}[J^*,J,\V] = \int_c P_G(\tau_{1},\tau_0) \  e^{- \AA_{Tot}^{(M)}}, \nn \\
&&\AA_{Tot}= \AA^{(M)}_0+\AA^{(M)}_S+\AA^{(M)}_t+\AA^{(M)}_J. 
\llabel{ZM}
 \eeq
We detail the various contributions next; the free Fermi term is given by
\barray
 \frac{1}{\Delta \tau} \Ag_0^{(M)}= \sum_{j=0}^{M-1} \left[ c^*_{i \si}(\tau_{j+1}) \delta_{\tau_j} c_{i \si}(\tau_j) 
- \chem \; n_{j \si}(\tau_j) \right], 
\earray
with the finite difference operator $\delta_{\tau_j}$ defined through
\barray
 \delta_{\tau_j} F(\tau_j) \equiv  \frac{1}{\Delta \tau} \left\{ F(\tau_{j+1})- F(\tau_{j})\right\}.
\earray
As $M\to \infty$,  we note that  $\delta_{\tau_j}$  reduces to the derivative operator $ \partial_\tau$, and   we obtain  the integral 
$\Ag_0 = \int_0^\beta d\tau \;  c^*_{i \si}(\tau) (\partial_\tau - \chem)  c_{i \si}(\tau)$, and in that limit
$Z^{(M)} \to Z[J^*,J,\V] $.
The source term $\AA_S^{(M)}(\tau_{j+1},\tau_j)$  obtained from \dispop{asource} is given by 
\barray
&&\AA^{(M)}_S=  \sum_i   \left[  \widetilde{c~}^*_{i \si}(\tau_{j+1})\ J_{i \si}(\tau_{j+1}) + J^*_{i \si}(\tau_{j+1}) c_{i \si}(\tau_j)\right] \nn \\
&& + \left[\V_{i}^{\si' \si}(\tau_{j+1}) c^*_{i \si'}(\tau_{j+1}) c_{i \si}(\tau_j) \right]. \llabel{asource2}
\earray
As in \dispop{asource}, the projected variable with  a hat appears in the creation operator and nowhere else in this expression. The Hamiltonian \disp{heff} gives rise to  two parts of the action.
The hopping term is given   by
\beq
\AA_t^{(M)}= \Delta \tau \sum_j T_{eff}(\tau_j) \to \int_0^\beta d\tau \ T_{eff}(\tau),
\eeq
with $T_{eff}$ from \disp{teff1} or \disp{teff2}: 
\beq
T^{Sym}_{eff}(\tau_j)&=&- \sum_{l m \si} t_{lm} \ c^*_{l \si}(\tau_{j+1}) c_{m \si}(\tau_{j}) \times \nn \\ &&\left( 1- n_{l \sib}(\tau_{j}) - n_{m \sib}(\tau_{j}) \right),\llabel{sym}\\
T^{Min}_{eff}(\tau_j)&=&- \sum_{l m \si} t_{lm} \ c^*_{l \si}(\tau_{j+1}) c_{m \si}(\tau_{j}) \times \nn \\ &&\left( 1- n_{l \sib}(\tau_{j})  \right),\llabel{minimal2}
\eeq
where \dispop{sym} corresponds to the symmetrized theory of \dispop{rephats} and \dispop{minimal2} to the minimal version of \dispop{rephats2}. The exchange part of the action is given by
\beq
&&\AA_J^{(M)}=\Delta \tau \sum_j H_J(\tau_j) \to \int_0^\beta d\tau \ H_J(\tau),\nn \\
&&H_{J}(\tau_j)\equiv   - \frac{1}{4} \sum_{l m  }  \ J_{l m}  \ {\si_1 \si_2} \times  \nn \\
&&  c^*_{l \si_1}(\tau_{j+1})  c^*_{m \sib_1}(\tau_{j+1}) c_{m \sib_2}(\tau_{j}) c_{l \si_2}(\tau_{j}). \llabel{eq321}
\eeq
Where possible  we    simplify the notation by dropping the superscript $M$; most   expressions provide sufficient context for this and there should be no confusion.  Thus we will write $ \G^{(M)}_{\si \si'}(a \tau_i, b \tau_f)\to \G_{\si \si'}(a \tau_i, b \tau_f) $ and $Z^{(M)} \to Z$ etc below.
When no confusion is likely we will refer to $Z[J^*,J,\V]$ as simply $Z$, and also abbreviate terms such as 
$\Ham_{eff}(\tau_{j+1},\tau_j)$  to $\Ham_{eff}(\tau_j)$ or even more simply to $ \Ham_{eff}$.
\disp{ZM} is almost in the form of a canonical partition function for unprojected electrons, but with an important difference. The  extra term in the integration measure is the Gutzwiller projector written in  Grassman variables. These variables arise  {\em at the initial and next time instant only} and the rest of the time variables have only the standard measure of unity. Explicitly we find
\beq 
P_G(\tau_{1},\tau_0) \equiv \prod_{i=1}^{N_s} \left( 1- c^*_{i \up}(\tau_1)c_{i \up}(\tau_0) c^*_{i \dn}(\tau_1)c_{i \dn}(\tau_0), \right) \nn \\ \llabel{gutzwiller}
\eeq
it has all creation (destruction) operators at $j=1$ ($j=0$), and  $N_s$ is the number of sites. 
In Appendix \ref{pauli} , we summarize  the Pauli principle and Gutzwiller identities obeyed by the present coherent state representation, these represent an important aspect of the strong correlation problem.
We will also recycle  the notation of \dispop{angles}   for the average in this distribution of any function $Q$ of the Grassman variables:
\beq \lll Q  \rrr & =& \frac{\llv Q \rrv}{Z} , \;\; \nn \\
\mbox{with} \;\;\llv Q \rrv &=&  \int_c P_G(\tau_{1},\tau_0) e^{- \AA} \ Q,  \llabel{vertbars}  \eeq
 a useful abbreviation \dispop{vertbars}, and drop the superscript $(M)$. This representation of the path integral with a constraining projection factor at only the initial time has a resemblance to the that in the canonical quantization of the electromagnetic field in the  temporal  gauge \cite{dirac,creutz}, as already noted in the introduction.  The Green's functions follow from \disp{eq23} using $\delta/\delta J(\tau_j) \to \frac{1}{(\Delta \tau)} d/ d J(\tau_j)$ \refdisp{simplify1}: 
\barray
\G_{\si_i \si_f}(i \tau_i, f \tau_f) 
&=& \frac{1}{Z} \llv \widetilde{c~}^*_{f \si_f}(\tau_{f})  \  c_{i \si}(\tau_i) \llv. \llabel{eq42}
\earray

\subsection{Equations of motion from path integral representation \label{path2}}
In this section we obtain the Schwinger equations of motion  of ECFL (see \refdisp{Monster} and especially Appendix~(\ref{Minimal}) \disp{EOM-42} ), directly from the path integral representation given above thus providing  a non trivial check on the representation.
To obtain \disp{EOM-42}, we initially set the Fermionic sources to zero,  the Bosonic sources are turned off at the very end. The equations of motion are most easily found using a  Grassman integration  identity:
\beq
&&\int_c P_G(\tau_{1},\tau_0)   \frac{\delta}{\delta c^*_{i \si_i}(\tau_{i+1})} \left[ \widetilde{c~}^*_{f \si_f}(\tau_{f+1}) \   e^{- \AA_{Tot}} \right] = 0, \nn \\
&&\llabel{eom1}
 \eeq
This identity is a straight forward  generalization of  the  theorem on vanishing of a total derivative \cite{Zinnjustin},    including  a non trivial measure $P_G$ \dispop{gutzwiller}
where the time arguments  are greater than all time arguments in \dispop{gutzwiller},  i.e. $i,f\geq1$. It is proved  by the usual logic for Grassman variables;   the derivative $\frac{\delta}{\delta c^*_{i \si_i}(\tau_{i+1})}$  is in addition to an integration over $ c^*_{i \si_i}(\tau_{i+1})$ contained in the overall integration.   We next recall that the highest possible degree of a  polynomial in any Grassman variable is  unity.   The above expression vanishes upon further  noting that Grassman integration and  Grassman differentiation are identical.
 The same identity is valid if we replace  $\widetilde{c~}^*_{f \si_f}(\tau_{f+1})$ by any other allowed  Grassman variable $U$, subject to the double occupancy restriction, and similarly with $V$ (see \refdisp{caution}).
 In  summary,  an abstract equation of motion, following from $\int P_G \frac{\delta }{\delta V} (U e^{-A_{Tot}}) =0$ and Fermionic $U,V$  reads
\beq
\llv \frac{\delta U}{\delta V} \llv + \llv U \frac{ \delta \AA_{S} }{\delta V} \llv + \llv U \frac{ \delta \AA_{0} }{\delta V} \llv+\llv U \frac{ \delta \AA_{t} }{\delta V} \llv+ \llv U \frac{ \delta \AA_{J} }{\delta V} \llv=0. \nn \\ \label{funda}
\eeq

\subsection{Equation for $\G_{\si_i \si_f}(i,f)$}
Our first task is to find an equation for the Green's function \cite{simplify1}- we use   \dispop{funda}  with $U= \widetilde{c~}^*_{f \si_f}(\tau_{f})$ and $V= c^*_{i \si_i}(\tau_{i})$.
We compute the various pieces of \dispop{funda} next.

Denoting 
 \beq \widehat{\gamma}_{\si_i \si_f }(i) \equiv {\si_i \si_f}   {c}^*_{i \sib_i}(\tau_{i+1}) {c~}_{i \sib_f}(\tau_{i}), 
 \llabel{gammadef} \eeq
 and using  the convention that repeated spin indices are summed over,  we obtain the  first  result:
  \beq 
  \frac{\delta}{\delta c^*_{i \si_i}(\tau_{i+1})}  \widetilde{c~}^*_{f \si_f}(\tau_{f+1}) =\delta_{\tau_i \tau_f}\delta_{if}\left\{ \delta_{\si_i \si_f} - \widehat{\gamma}_{\si_i \si_f }(i)\right\}.\; ~
   \llabel{derdef}
  \eeq

 We obtain 
\beq
\frac{1}{\Delta \tau}\frac{\delta \AA_0}{\delta c^*_{i \si_i}(\tau_{i+1})}& = & \delta_{\tau_{i}}  c_{i \si_i}(\tau_{i}) - \chem \; c_{i \si_i}(\tau_{i}),\nn \\
\eeq  
\beq
&&\frac{1}{\Delta \tau}\frac{\delta \AA_S}{\delta c^*_{i \si_i}(\tau_{i+1})}= \V_i^{\si_i \si_j}(\tau_{i+1}) \ c_{i \si_j}(\tau_i) \nn \\
&& + \left\{ \delta_{\si_i \si_j} - \widehat{\gamma}_{\si_i \si_j }(i)\right\} J_{i \si_j}(\tau_{i+1}),
\eeq
\beq
&&\frac{1}{\Delta \tau}\frac{\delta \AA^{Sym}_t}{\delta c^*_{i \si_i}(\tau_{i+1})}= - \nn t_{ij} c_{j \si_i}(\tau_i)\\
&&+ t_{ij} \left[ \hat{\gamma}_{\si_i \si_j}(i \tau_i) \ c_{j \si_j}(\tau_i) + c^*_{j \sib_i}c_{j \sib_i} c_{j \si_i} + c^*_{j \sib_i}c_{i \sib_i} c_{i \si_i}   \right], \nn  \\
&&\frac{1}{\Delta \tau}\frac{\delta \AA^{Min}_t}{\delta c^*_{i \si_i}(\tau_{i+1})}= - \nn t_{ij} c_{j \si_i}(\tau_i)\\
&&+ t_{ij} \hat{\gamma}_{\si_i \si_j}(i \tau_i) \ c_{j \si_j}(\tau_i) \label{tder}
\eeq  
\beq
&&\frac{1}{\Delta \tau}\frac{\delta \AA_J}{\delta c^*_{i \si_i}(\tau_{i+1})}=- \frac{1}{2}  J_{i j}   {\si_i \si_j}\,   \nn \\
&&c^*_{j \sib_i}(\tau_{i+1}) c_{j \sib_j}(\tau_{i}) c_{i \si_j}(\tau_{i}), \label{Jder}
\eeq  
We combine the two terms as:
\barray
\frac{1}{\Delta \tau}\frac{\delta (\AA_t+ \AA_J)}{\delta c^*_{i \si_i}(\tau_{i+1})}= 
  -\sum_j t_{ij} c_{j \si_i}(\tau_i)  + A_{i \si_i}(\tau_{i+1},\tau_i) \llabel{eq421},\nn \\
\earray
 the first   (linear)  term in Fermions is  separated out in this expression, and
 $A_{i \si_i}$, detailed below in \disp{eq43}, is obtained by  combining all  the {\em three Fermion} contributions  in \disp{tder} and \disp{Jder}. 
 In the minimal  case we get from \disp{minimal2}, \disp{eq321}, \disp{gammadef} and \disp{derdef}
 \barray
 A^{Min}_{i \si_i} =   t_{ij} \widehat{\gamma}_{\si_i \si_j}(i \tau_i) \ c_{j \si_j}(\tau_i) 
  - \frac{1}{2}    J_{i j} \widehat{\gamma}_{\si_i \si_j}(j \tau_i) c_{i \si_j}(\tau_{i}), \nn \\\llabel{eq43}
 \earray
 in agreement with Eq.~(22) of \refdisp{Monster},
 and the symmetrized case is obtained in a similar way. Combining these (with $J \to 0$)   we get the EOM in discrete time space:
 \begin{widetext}
\barray
&& \left[ \left\{\chem - \delta_{\tau_i} -  \V_i^{\si_i \si_j}(\tau_{i+1}) \right\} \delta_{i,j}+  t_{ij} \right]     \llv  \widetilde{c~}^*_{{f} \si_f}(\tau_{f+1})  c_{j \si_i}(\tau_{i})
 \llv  
 -\llv  \widetilde{c~}^*_{{f} \si_f}(\tau_{f+1}) A_{i \si_i}(\tau_{i+1},\tau_i)
 \llv =\delta_{if}  \llv  ( \delta_{\si_i \si_f} - \widehat{\gamma}_{\si_i \si_f }(i) ) \llv \ \frac{ \delta_{\tau_i,\tau_{f}}}{\Delta \tau} \nn \\  \llabel{eq45}
\earray
\end{widetext}
We next take the continuum limit in time $\tau_i$;  with  $\Delta \tau \to 0$,   and using $ \frac{ \delta_{\tau_i,\tau_{f}}}{\Delta \tau} \to \delta(\tau_i-\tau_f)$, and using  the non interacting  Fermi  Green's function from \dispop{gnon},
 and implementing the basic Schwinger identity for  representing higher order correlation functions as source derivatives:
 \beq
  \llv  \widetilde{c~}^*_{{f} \si_f}(\tau_{f}) A_{i \si_i}(\tau_i)\llv = \hat{X}_{ \si_i \si_j}(i \tau_i, j \tau_j) \llv\widetilde{c~}^*_{{f} \si_f}(\tau_{f})   c_{j \si_j}(\tau_j)  \llv\nn \\ \llabel{xdef}
 \eeq
 where $\hat{X}$ is a functional derivative operator defined more completely below in \dispop{xopdef}. With this preparation
 we can rewrite \disp{eq45}  as
 \beq
&& \left(  \GHI_{0, \si_i,  \si_j}(i \tau_i,  j \tau_j) -  \hat{X}_{ \si_i \si_j}(i \tau_i, j \tau_j)\right)\ \llv \widetilde{c~}^*_{{f} \si_f}(\tau_{f})   c_{j \si_j}(\tau_j)  \llv \nn \\
 &&=\delta_{if}  \llv  ( \delta_{\si_i \si_f} - \widehat{\gamma}_{\si_i \si_f }(i) ) \llv \delta(\tau_i-\tau_f).   \llabel{eq46}
 \eeq
 Here and elsewhere since $\tau_j$ repeats in product, it  is assumed to be integrated between $0 \leq \tau_j \leq \beta$, this rule is analogous to the spin index summation rule. We next divide  by $Z$, use \dispop{vertbars} to define the Green's function, and also define
 \beq
  {Y_1}_{ \si_i \si_j}(i \tau_i, j \tau_j) =\frac{1}{Z}  (\hat{X}_{ \si_i \si_j}(i \tau_i, j \tau_j)  Z), \llabel{y1def}
 \eeq
 to rewrite \dispop{eq46} in the same form as \disp{EOM-42}
 \beq
  \left(  \GHI_{0, \si_i,  \si_j}(i \tau_i,  j \tau_j) -  \hat{X}_{ \si_i \si_j}(i \tau_i, j \tau_j)-  {Y_1}_{ \si_i \si_j}(i \tau_i, j \tau_j)\right)\times \nn \\
   \G_{\si_i \si_j}( i \tau_i, j \tau_j) = \delta_{if} \delta(\tau_i-\tau_f)  \left[ \delta_{\si_i \si_f} - \gamma_{\si_i \si_f }(i) \right],  \nn \\
 \eeq
where
\beq
\gamma_{\si_i \si_f }(i)\equiv \lll \widehat{\gamma}_{\si_i \si_f }(i) \rrr. \label{gamma}
\eeq
This is readily seen to be identical to the direct definition given before in \dispop{gamma-def}.
We next use  $\bh{i} \equiv (i,\tau_i,\si_i)$ as an abbreviation for the (space, time, spin) indices, and use the repeated index summation convention. Here summation stands for spin and  spatial sums,  and temporal integrals in the standard intervals. 
With this we can write $  \hat{X}_{ \si_i \si_j}(i \tau_i, j \tau_j)\leftrightarrow \hat{X}_{\bh{i} \bh{j}}$, and similarly for $\GH^{-1}_0$, $\G$ and $Y_1$. 
The variable \dispop{gamma} is local and needs the extra definition $\gamma_{\si_i \si_f}(i \tau_i) \delta_{if} \delta(\tau_i-\tau_f) \leftrightarrow \gamma_{\bh{i} \bh{f}}$ and also denote $ \delta_{if}    \delta_{\si_i \si_f} \delta(\tau_i-\tau_f) \leftrightarrow \delta_{\bh{i} \bh{f}}$. With these, the matrix product form of \disp{eq46} reads:
 \beq
  \left(  \GHI_{0, \bh{i}  \bh{j} } -  \hat{X}_{\bh{i} , \bh{j} }-  {Y_1}_{\bh{i} , \bh{j} }\right) ~\G_{  \bh{j},\bh{f}} = (\delta_{\bh{i} \bh{f}} - \gamma_{\bh{i} \bh{f}}). \llabel{eom2} \nn \\
 \eeq
This is exactly the form of the Schwinger equation for the Green's function obtained 
from the continuous time Heisenberg equations of motion \disp{EOM-42}
in \cite{ECFL,Monster}, using the above abbreviation convention.

 In order to obtain an expression for $\hat{X}$, we note  a useful relationship involving the action on the partition functional \dispop{ZM} of the  operator  $D_{\si_i \si_j}(i) \equiv \si_i \si_j \delta/\delta \V_{i}^{\sib_i \sib_j} $ (from Eq~(39) of \refdisp{Monster}) 
 \beq
 D_{\si_i \si_j}(i)  Z[\V] = - \llv \hat{\gamma}_{\si_i \si_j}(i), \rrv  \label{gamma-origin}
  \eeq
so that:
\beq
&&\llv  \widetilde{c~}^*_{{f} \si_f}(\tau_{f+1}) A^{Min}_{i \si_i} \rrv=-  t_{ij} D_{\si_i \si_j}(i) \llv  \widetilde{c~}^*_{{f} \si_f}(\tau_{f+1}) c_{j \si_j}(\tau_i)\llv \nn \\
&&+ \frac{1}{2}  J_{ij} D_{\si_i \si_j}(j) \llv  \widetilde{c~}^*_{{f} \si_f}(\tau_{f+1}) c_{i \si_j}(\tau_i)\llv,
\eeq
and  comparing with \disp{xdef} we conclude
\beq
&&\hat{X}_{\si_i \si_j}(i \tau_i, j \tau_j)=\delta(\tau_i-\tau_j) \times \nn \\
&&( - t_{ij} D_{\si_i \si_j}(i)+ \delta_{ij} \sum_k \frac{1}{2} J_{ik} D_{\si_i \si_j}(k \tau_i ) ),\llabel{xopdef}
\eeq
where the derivative $ D_{\si_i \si_j}(k \tau_i )$ is  at spatial site $k$ and time $\tau_i$. The corresponding $Y_{1}$ (with a similar convention as above)  in \disp{y1def} is 
\beq
&&{Y_1}_{\si_i \si_j}(i \tau_i, j \tau_j)=- \delta(\tau_i-\tau_j) \times \nn \\
&&( - t_{ij} \gamma_{\si_i \si_j}(i)+ \delta_{ij} \sum_k \frac{1}{2} J_{ik} \gamma_{\si_i \si_j}(k \tau_i) ).\llabel{ydef}
\eeq
Analogous expressions for the symmetrized case,
for $A_i$, $\hat{X}$ and $Y_1$  parallel to \disp{xopdef}, \disp{ydef} and \disp{eq43}, 
 can be obtained  by using the symmetrized version (top line) of \disp{tder}. These
expression agrees with the  Eq.~(43) of \refdisp{Monster}, and their minimal version obtained after dropping the second and fourth term. 
We have thus verified that the exact equations of motion are obtained from the path integral representation outlined here,  constituting a non trivial check on the formalism.

\section{Conclusions \label{conclusions}}
In this work we have presented a simpler method to obtain the ECFL theory that complements the Schwinger method used earlier. This new method brings an important analogy to the Dyson-Maleev theory to attention, and this connection helps us to get a different perspective on the main results of ECFL, in particular the novel non-Dysonian representation of the  Greens function is placed on a firm foundation. The path integral method is used to set up an alternate quantum field theory with a non Hermitean Hamiltonian, and it is proven to be valid by reproducing the Schwinger equations of motion. 

We draw particular attention to the scaling result for the spectral function \disp{scaling} and \disp{scaling2} in Section~(\ref{2D}). Here the low energy spectral function is shown to satisfy a simple relation  involving the hole density that throws light on the  ever shrinking regime of validity of the Landau Fermi liquid, as we approach the insulating state. Finally the discussion of the alternate ways to analyze the ARPES line shapes discussed in Section~(\ref{2E}) should be of interest to the ARPES community, as also the discussion of the electronic origin of a kink in the EDC energy dispersion.

\section{Acknowledgements}
This work was supported by DOE under Grant No. FG02- 06ER46319.
I thank T. Banks and O. Narayan for helpful discussions on the    path integral representation. I am   grateful to P. W. Anderson,  P. Coleman, A. Georges, G. H. Gweon, A. Hewson, P.D. Johnson, G. Kotliar, H. R. Krishnamurthy,  E. Perepelitsky and T. M. Rice for helpful comments and valuable discussions.  

\appendices
\section{Summary of the minimal theory and its  Schwinger equations of Motion \label{Minimal}}
In order to make the discussions reasonably self contained, we provide a brief discussion of the minimal equations of motion for the Green's function. These are obtained through the usual Schwinger method used in \refdisp{ECFL} and in \refdisp{Monster}. These equations are a subset of the ones given in \refdisp{Monster}, and can be obtained by omitting certain extra terms therein, which were added to satisfy a symmetry property. We term these equations are {\em the minimal theory}, since no terms are added or dropped, and the expressions are not reducible by any other argument.
  We also indicate the generalization to include the parameter $\lambda$ in these equations, to facilitate comparing with the equations in this work. 

Using the Hamiltonian \dispop{hamiltonian} we note 
  the important commutator (given in \refdisp{Monster}): 
\barray
&&[ H_{tJ},\X{i}{0 \si_{i}}]  =   \sum_j t_{ij} \X{j}{0 \si_i}  
 + \chem \X{i}{0 \si_i}  \nn \\
&& - \sum_{j \si_j} t_{ij} (\si_i \si_j)  \X{i}{\sib_i \sib_j} \ \X{j}{0 \si_j} + \frac{1}{2} \sum_{j \neq i} J_{ij} \ (\si_i \si_j)   \X{j}{\sib_i \sib_j}   \X{i}{0 \si_j}. \nn \\ \label{a} 
\earray
Temporarily ignoring the Fermionic sources:
\beq
[\A_S(i \tau_i),\X{i}{0 \si_{i}}]= -\V_i^{\si_i\si_j} \X{i}{0 \si_j}, \label{aa}
\eeq
and combining with the Heisenberg equation of motion, we see that the Green's function satisfies the EOM
\beq
&&\partial_{\tau_i} \G_{\si_i \si_f}(i , f)= - \delta(\tau_i-\tau_f) \delta_{ij} (1- \gamma_{\si_i \si_f}(i \tau_i) ) \nn \\
&&- \langle T_\tau \left( e^{- \A_S} [H_{tJ}+ \A_S(i,\tau_i), \X{i}{0 \si_i}(\tau_i)] \ \X{f}{\si_f 0}(\tau_f) \right) \rangle \nn \\
\eeq
where the local Green's function 
 \beq\gamma_{\si_a \si_b}(i \tau_i) = \si_a \si_b \G_{\sib_b \sib_a}(i \tau_i^-, i \tau_i). 
  \label{gamma-def}
 \eeq
Substituting and using the  Fermi gas ( i.e. free) Green's function:
\beq
 &&\GHI_{0, \si_i,  \si_j}(i \tau_i,  j \tau_j)=\nn \\
&&  \left\{  \delta_{\si_i \si_j} \left[\delta_{ij} (\chem- \partial_{\tau_i}) + t_{ij} \right]  - \delta_{ij} \V_i^{\si_i \si_j}(\tau_i)\right\} \delta(\tau_i-\tau_j), \nn \\ \label{gnon} 
 \eeq
we obtain (using the repeated index summation and integration convention of \refdisp{Monster})
\beq
&&\GHI_{0, \si_i,  \si_j}(i \tau_i,  j \tau_j) \G_{\si_j \si_f}(j \tau_j , f \tau_f)= \nn \\
&&  \delta(\tau_i-\tau_f) \delta_{ij} (1- \gamma_{\si_i \si_f}(i \tau_i) ) \nn \\
&& - \sum_{j \si_j} t_{ij} (\si_i \si_j) \   \langle T_\tau \left(\X{i}{\sib_i \sib_j}(\tau_i)  \X{j}{0 \si_j}(\tau_i)  \ \X{f}{\si_f 0}(\tau_f) \right)\rangle \   \nn \\
&&+ \frac{1}{2} \sum_{j \si_j} J_{ij}  (\si_i \si_j) \langle T_\tau \left(   \X{j}{\sib_i \sib_j}(\tau_i)   \X{i}{0 \si_j}(\tau_i)  \X{f}{\si_f 0}(\tau_f) \right)\rangle \ .\nn \\ \label{EOM-421} 
\eeq
 We next express the higher order Green's function in terms of the derivatives of the lower one to obtain the Schwinger EOM:
 \beq
  \left(  \GHI_{0, \si_i,  \si_j}(i \tau_i,  j \tau_j) -  \hat{X}_{ \si_i \si_j}(i \tau_i, j \tau_j)-  {Y_1}_{ \si_i \si_j}(i \tau_i, j \tau_j)\right) \nn \\
  \times \G_{\si_j \si_f}( j \tau_j, f \tau_f) = \delta_{if} \delta(\tau_i-\tau_f)  \left( \delta_{\si_i \si_f} - \gamma_{\si_i \si_f }(i \tau_i) \right),  \nn \\ \label{EOM-42}
 \eeq
 where
we used the functional derivative operator
\barray
 {D}_{\si_i \si_j}(i \tau_i)& = &  \si_i \si_j { \frac{\delta}{\delta \V_i^{\sib_i \sib_j}(\tau_i)}}  \label{def-der}
 \earray
and the composite derivative operator
\beq
&&\hat{X}_{\si_i \si_j}(i \tau_i, j \tau_j)=\delta(\tau_i-\tau_j) \times \nn \\
&&\left( - t_{ij} D_{\si_i \si_j}(i \tau_i)+ \delta_{ij} \sum_k \frac{1}{2} J_{ik} D_{\si_i \si_j}(k \tau_i ) \right),\llabel{xopdef2}
\eeq
where the derivative $ D_{\si_i \si_j}(k \tau_i )$ is  at spatial site $k$ and time $\tau_i$. The corresponding $Y_{1}$ (with a similar convention as above)  in \disp{y1def} is 
\beq
&&{Y_1}_{\si_i \si_j}(i \tau_i, j \tau_j)=- \delta(\tau_i-\tau_j) \times \nn \\
&&\left( - t_{ij} \gamma_{\si_i \si_j}(i \tau_i)+ \delta_{ij} \sum_k \frac{1}{2} J_{ik} \gamma_{\si_i \si_j}(k \tau_i) \right).\llabel{ydef2}
\eeq
  Eqs.~[(\ref{EOM-42}) , (\ref{xopdef2}), and (\ref{ydef2})] define the minimal theory. For reference we note that \refdisp{Monster} gives these equations, and  goes on to  add terms that account for the symmetrized theory with a Hermitean $H_{eff}$.  We also note that the equation   \disp{EOM-42} can be generalized to include the $\lambda$ parameter  by scaling $\hat{X}_{\si_i \si_j}, Y_{i \si_i \si_j}, \gamma_{\si_i \si_j} \to \lambda \hat{X}_{\si_i \si_j}, \lambda Y_{i \si_i \si_j}, \lambda \gamma_{\si_i \si_j}$.

% \end{widetext}

\section{Coherent State Definitions \llabel{coherent-states}}
We use standard anticommuting Grassman variables \cite{Zinnjustin} to represent the canonical Fermions $\ch{}$ and $\chd{}$ for each spin and site. In brief we note the anticommuting property
 $\left\{c_i, c^*_j \right\}=0=\left\{c_i, c_j \right\}=\left\{c^*_i, c^*_j \right\}= \left\{c_i, \chd{j} \right\}$.
 Suppressing $j$ and spin  index the Fermi coherent states are given as usual by:
\barray
|c\rangle &=& e^{- c \ \chd{}} | vac \rangle = \left( 1 - c \ \chd{} \right) \ |vac \rangle\nn \\
\langle c | &=& \langle vac | e^{-  \ch{} \ c^* }  = \langle vac | \left( 1 -  \ \ch{} \ c^* \right) \ \nn \\
\langle c | c' \rangle &= & 1 + c^* c' = e^{ c^* c'}, 
 \earray
 where $|vac\rangle$ is the vacuum state.
 We use the abbreviation to denote coherent state integrals:
\beq
\int_c = \int d c^* \ dc.
\eeq
  The basic integrals are 
\barray
\int_c \left( 1, c^*,c, c c^*\right) &=& \left( 0,0,0,1\right) \nn \\
\int_c e^{- c^* c} &=& 1
\earray
The completeness relation reads:
\barray
\int_c e^{- c^* c} \  | c \rangle \langle c| = |vac \rangle \langle vac| +\chd{}  |vac \rangle \langle vac| \ch{} \equiv \iden,
\earray 
and the trace over Fermionic  variables is given by: 
\barray
\tr A = \int_c \ e^{-c^* c} \ \langle -c | A | c \rangle.
\earray

\section{Path integral representation \llabel{pathint}}
We now introduce path integrals to represent  \disp{angles2} leading to \dispop{ZM}. Towards this end let us write
\barray
\beta \hat{H}_{eff} + \A_S & = & \int_0^\beta \Hamc(\tau) d \tau \nn \\
\Hamc(\tau)&\equiv& \hat{H}_{eff} + \sum_i \A_S(i, \tau).
\earray
The integral is represented by a finite sum over $M$ intervals, and the limit $M\to \infty$ taken at the end, thus
\barray
\int_0^\beta \Hamc(\tau) d \tau&\to& \lim_{M \to \infty} \Delta \tau \ \sum_{j=1,M} \Hamc(\tau_j) \earray
Where we defined
\barray
&&\tau_j = \Delta \tau \times j =\frac{ j \ \beta}{M},\nn \\
&&\Delta \tau = \frac{\beta}{M},\;\;
j= 1, M.
\earray
Thus with $\Hamc(j) \equiv \Hamc(\tau_j)$ arranged  to be in normal ordered form (creation operators to the left of the destruction operators)  we write Trotters formula for the exponential
\begin{widetext}
\barray
Z^{(M)} &=& \int_{c(0)} e^{- c_{i \si}^*(0) c_{i \si}(0)} \ \langle - c(0)| e^{- \Delta \tau    \Hamc(\tau_M) } \ e^{- \Delta \tau    \Hamc(\tau_{M-1}) }\ldots   \ e^{- \Delta \tau    \Hamc(\tau_2) }  e^{- \Delta \tau    \Hamc(\tau_1) } \ \hat{P}_G    |c(0)\rangle  \nn \\
 &=& \int_{c} e^{- \sum_{j=1}^{M} c_{i \si}^*(j) c_{i \si}(j)} \ \langle  c(M)| e^{- \Delta \tau    \Hamc(\tau_M) } |c(M-1) \rangle \ldots   \ |c(2) \rangle \langle c(2)|  e^{- \Delta \tau    \Hamc(\tau_2) }|c(1) \rangle \langle c(1)|   e^{- \Delta \tau    \Hamc(\tau_1) } \ \hat{P}_G    |c(0)\rangle.  \nn \\ \llabel{eq25}
\earray
  Anti periodic boundary conditions  are used: $c(\tau_M) = -c(\tau_0)$ and we set at each time slice $\tau_j$ the coherent state $|c(j)\rangle = \prod_{i \si} |c_{i \si}(\tau_j) \rangle$ as a global product over all sites and both spins, and 
the symbol $\int_c$ represents integration over all the sites spins and time slices. The site index $i$ and spin $\si$ are implicitly summed over. Recall that  $\hat{P}_G$ is brought to the extreme right of the product. We calculate as usual:
\barray
\langle c(\tau_{j+1})| c(\tau_j) \rangle &=& e^{c^*(\tau_{j+1})c(\tau_j)} \nn \\
\langle c(\tau_{j+1})| e^{- \Delta \tau \Hamc({\tau_{j+1}})} | c(\tau_j) \rangle &\equiv & e^{c^*(\tau_{j+1})c(\tau_j) - \Delta \tau \Ham(\tau_{j+1}, \tau_j)}  + O(\Delta \tau^2)  \nn \\
\Ham(\tau_{j+1}, \tau_j)&\equiv& \frac{ \langle c(\tau_{j+1})| \Hamc(\tau_{\tau_{j+1}}) | c(\tau_j) \rangle }{\langle c(\tau_{j+1}) | c(\tau_j) \rangle }. 
\earray

The last term  needs careful attention, we note
\barray
 \langle c(\tau_1)|   e^{- \Delta \tau    \Hamc(\tau_1) } \ \hat{P}_G    |c(\tau_0)\rangle&=& \langle c(\tau_1)|  ( {1 - \Delta \tau    \Hamc(\tau_1) }) \ \hat{P}_G    |c(\tau_0)\rangle +O(\Delta \tau^2) \nn \\
 &=&  \langle c(\tau_1)| c(\tau_0) \rangle  ( {1 - \Delta \tau    \Ham(\tau_1,\tau_0) }) \ {P}_G(\tau_1,\tau_0)   +O(\Delta \tau^2)   \nn \\
 &=& \langle c(\tau_1)| c(\tau_0) \rangle e^{ - \Delta \tau    \Ham(\tau_1,\tau_0) } \ {P}_G(\tau_1,\tau_0)   +O(\Delta \tau^2),   
\earray
\end{widetext}
where \disp{gutzwiller}  details  the expression for  ${P}_G(\tau_1,\tau_0)$,  it contains variables    at the initial and next time instant only. Combining all terms, we get the expression \disp{ZM}. We have thrown out terms of $O(\Delta\tau)^2$ in obtaining \disp{ZM}, and hence it is important to keep track of the Pauli principle identities, discussed above in \disp{pauli-1} and \disp{pauli-2}.
Note that for arbitrary $\tau$
\beq
\langle \tau | \chd{i \si} | \tau_j\rangle = c^*_{i \si}( \tau) e^{ c^*_{i \si}( \tau) c_{i \si}(\tau_j)} = - \frac{\delta}{\delta c_{i \si}(\tau_j)}   \langle \tau| \tau_j \rangle.
\eeq
In view of this relation  we  note the following mappings:
\barray
\langle \psi  | \chd{i \si}  | \tau_j\rangle & \to& - \frac{\delta}{\delta c_{i \si}(\tau_j)} \langle \psi   | \tau_j\rangle, \nn \\
\langle \psi  |\ch{i \si} | \tau_j\rangle & \to& { c_{i \si}(\tau_j)} \langle\psi  | \tau_j\rangle, \nn \\
 \langle  \tau_j  |  \chd{i \si} |\psi \rangle & \to& { c^*_{i \si}(\tau_j)} \langle   \tau_j | \psi \rangle, \nn \\
  \langle  \tau_j  |  \ch{i \si} |\psi \rangle & \to& \frac{\delta}{\delta c^*_{i \si}(\tau_j)} \langle   \tau_j | \psi \rangle, \label{dermaps}
\earray
Let us show how the commutation works here:
\beq
&&(\ch{}\chd{}+\chd{}\ch{})|c \rangle= (- \ch{} \frac{\delta}{\delta c}+ \chd{} c ) |c\rangle =  (  \frac{\delta}{\delta c} \ch{} -  c \chd{} ) |c\rangle \nn \\
&& = (  \frac{\delta}{\delta c} c +   c \frac{\delta}{\delta c}  ) |c\rangle =|c\rangle . 
\eeq

\section{Pauli  and Gutzwiller exclusion  identities \llabel{pauli}}
It is worth highlighting a few conventions about  \dispop{ZM} and related expressions. These are designed to retain some of the most important features of strongly interacting electrons {\em on a lattice}. In contrast,  in a theory of electrons in the continuum, these constraints are of no special consequence- since coincident spatial points have a measure of zero. 
We first discuss the Pauli principle related rules referring to the same spin spices, and then the Gutzwiller projection related rules relating to opposite spin species, these are operative when two electronic operators have coincident space and time coordinates.
   \begin{itemize}
  \item{(I)}
   When two coincident times in a product of operators {\em on the same lattice site}  and same spin arise, we follow the convention of immediate  evaluation of  the product.  By evaluation, we understand that the product of two similar Grassman variables  is set to zero, and for dissimilar Grassman variables (e.g. $c$ and $c^*$) at a  common time,  both of them are integrated out immediately. This leads to the   basic set of {\em  Pauli exclusion identities} at equal times as one easily verifies:
\barray
c_{i \si}(\tau_j)c^*_{i \si}(\tau_j)&\to& 1  \nn \\
c^*_{i \si}(\tau_j)c_{i \si}(\tau_j)&\to& -1 \nn\\
c_{i \si}(\tau_j)c_{i \si}(\tau_j)&\to& 0 \nn\\
c^*_{i \si}(\tau_j)c^*_{i \si}(\tau_j)&\to& 0. \llabel{pauli-1}
\earray

\item{(II)} We denote the number operator as
\beq
 n_{i \si}(\tau_j) \equiv c^*_{i \si}(\tau_{j+1}) \  c_{i \si}(\tau_j),
\eeq
where we observe  that  the $c^*$  has the immediately later time argument than that of $c$, this comes about from representing $\langle \ j+1| C^\dagger C |j\rangle= c^*(\tau_{j+1}) c(\tau_j) \times \langle j+1|j\rangle$. Using this we will verify the  {\em second  set} of {
  Pauli exclusion identities}
\barray
n_{i \si}(\tau_{j+1})n_{i \si}(\tau_j) &=& n_{i \si}(\tau_j) \nn\\
c^*_{i \si}(\tau_{j+1}) n_{i \si}(\tau_{j})&=& 0 \nn \\
 n_{i \si}(\tau_{j})c^*_{i \si}(\tau_{j})&=& c^*_{i \si}(\tau_{j+1}) \nn \\
 n_{i \si}(\tau_{j}) c_{i \si}(\tau_{j})&=& 0 \nn \\
  c_{i \si}(\tau_{j})n_{i \si}(\tau_{j-1})&=& c_{i \si}(\tau_{j-1}) \llabel{pauli-2}
\earray
\itemize{(III)}
We next  obtain the important {\em Gutzwiller exclusion identity}.
  Calling   the  $i^{th}$ term in the product  \dispop{gutzwiller}   as $P^{(i)}_G(\tau_{1},\tau_0)$;   we see that
\barray
&&n_{i \up}(\tau_1)n_{i \dn}(\tau_1) P^{(i)}_G(\tau_{1},\tau_0)=  n_{i \up}(\tau_1)n_{i \dn}(\tau_1) \nn \\
&& - c^*_{i \up}(\tau_{2})c_{i \up}(\tau_{0})c^*_{i \dn}(\tau_{2})c_{i \dn}(\tau_{0}) \sim 0. 
\llabel{gutzvanish}
\earray
 The last line follows upon  expanding $\tau_2(\equiv \tau_1+ \Delta\tau)$ about $\tau_1$.  The  assumption  that terms of $O(\Delta \tau)$ are negligible is  implicit in the entire path integral formulation. This shows that the double occupancy type terms
$n_{i \up}(\tau_1)n_{i \dn}(\tau_1)$ that occur at { any  site} lead to vanishing contribution, thus enforcing Gutzwiller projection.
{ We can extend this argument to other times $\tau_j \geq \tau_1$:
\beq
c_{i \up}(\tau_j)c_{i \dn}(\tau_j) \ldots P^{(i)}_G(\tau_{1},\tau_0)=0,  \nn \\ \llabel{gutzvanish2}
\eeq
where the dots indicate contributions from intermediate times.
These contributions,  after Grassman integration  over  the terms at intermediate times,  must necessarily  end up  with $\ldots c_{i \up}(\tau_1)c_{i \dn}(\tau_1)  P^{(i)}_G(\tau_{1},\tau_0)$.
Expanding this factor
 leads to $c_{i \up}(\tau_1)c_{i \dn}(\tau_1)-c_{i \up}(\tau_0)c_{i \dn}(\tau_0) $, and therefore vanishes to $O(\Delta \tau)$, as in  the argument in \dispop{gutzvanish}.
 }
\end{itemize}

\section{Interpreting the caparison factor in the Schwinger  method \label{AE}} 
Within the Schwinger method, or the related  path integral formulation given above, the decomposition of $\G$ is best done by rescaling  the source terms by a factor determined  through 
 a self consistent argument given next.   A convenient method is to work  in the presence of the Fermionic sources, which allows us to start with the non vanishing average of a Fermi operator:
\beq
\xi_{i \si_i}(\tau_i) \equiv \frac{1}{Z} \llv X_i^{0 \si_i}(\tau_i) \llv = \frac{1}{Z} \llv c_{i \si_i}(\tau_i) \llv, 
\eeq 
and further abbreviate $\xi_{i \si_i}(\tau_i) \leftrightarrow  \xi_{\bh{i}} $. A creation variable  average $\xi^*_{\bh{i}}\equiv \frac{1}{Z} \llv X_i^{ \si_i 0}(\tau_i) \llv$ is also useful.
 The  variable $\xi_{\bh{i}}$ satisfies the functional differential equation that we study next:
\beq
(\GH^{-1}_{0,\bh{i},\bh{j}} - \hat{X}_{\bh{i} \bh{j}} ) (Z \xi_{\bh{j}})=  Z  (\delta_{\bh{i}, \bh{k}}- \gamma_{\bh{i},\bh{k}}) J_{\bh{k}}, \label{eq59}
\eeq
or using \disp{y1def} 
\beq
(\GH^{-1}_{0,\bh{i},\bh{j}} - \hat{X}_{\bh{i} \bh{j}} - {Y_1}_{\bh{i} \bh{j}} ) ~ \xi_{\bh{j}} =    (\delta_{\bh{i}, \bh{k}}- \gamma_{\bh{i},\bh{k}}) J_{\bh{k}}. \label{eq60}
\eeq
This equation can be arrived at within the  path integral representation \disp{eq23}, by using a variant of \dispop{eom1} after omitting the Fermionic creation type variable in the square bracket; and  of course with a non vanishing Fermi source term.  Alternately  we can take the Heisenberg equations of motion in terms of the original expressions in terms of the $X$ operators  \disp{part1} and \disp{green}. he agreement between the two methods can be checked easily, and provides a strong check on the path integral formulation.

The Green's function is found from a variant of \disp{green}:
\beq
\G_{\bh{i} \bh{f}}- \xi^*_{\bh{f}}~ \xi_{\bh{i}} = \frac{\delta \xi_{\bh{i}}}{\delta J_{\bh{f}}}, \label{gdef2}  
\eeq
and on taking the limit $J\to 0, J^* \to 0$, all the single Fermi expectations $\xi, \xi^*$ vanish.  Taking the $J$ derivative of \dispop{gdef2}, we see that \dispop{eom2} follows, so we will work with this equation from here.

The main objective from this point onwards, is to cast \disp{eq60} or \disp{eom2} into a form where the expressions on right are in the canonical form, i.e. where  the time dependent $\gamma$ term is gotten rid of in favor of  a suitable constraint \cite{ECFL,ECQL,ECQL2}.   The  occurrence of the factor $\iden - \gamma$ multiplying the source $J$ in \disp{eq60} suggests that one should scale the source $J$ by  a suitable time dependent factor to obtain new sources $\I$. The factor can be adjusted self consistently, so as to extract  a canonical Green's function. Thus we scale
\beq
J_{\bh{i}}= \muin_{\bh{i} \bh{j}} \ \I_{\bh{j}}, \label{jscale}
\eeq 
so that 
\beq
\frac{\delta}{\delta J_{\bh{f}}} = \frac{\delta}{\delta \I_{\bh{k}}} \mu_{\bh{k} \bh{f}}, \label{jderscale}
\eeq
where $\muin$ is the  {\em matrix inverse} of $\mu$. These equations have inverses that are easily obtained.
The matrix $\mu$ is dependent on the Fermi sources only indirectly, and this dependence may be neglected since  the Fermi sources are turned off in the sequel. However it is
 allowed to be a functional of the Bosonic sources $\V$, thereby giving us considerable flexibility in defining it, we must also then be careful in locating it relative to the operator $\hat{X}$, since it contains derivatives with respect to $\V$.

In view of \disp{jderscale}  we  obtain the product relation
\barray
\G_{\bh{i} \bh{f}} & = & \GH_{\bh{i} \bh{k}} \ \mu_{\bh{k} \bh{f}} +  \xi^*_{\bh{f}}~ \xi_{\bh{i}}, \nn \\
\GH_{\bh{i} \bh{k}}&=&  \frac{\delta \xi_{\bh{i}}}{\delta \I_{\bh{k}}}.
\earray
The goal is to choose $\mu$ such that the so defined $\GH$ satisfies a canonical equation, i.e. the analog of \disp{eom2}, but without the $\gamma$ term on the right. For this purpose we can differentiate \disp{eq60} with the scaled source field $\I_{\bh{k}}$, taking care to observe the non commutation of $\I_{\bh{k}}$ with the derivative term $\hat{X}$. This process yields the equations:
\barray
&&(\GH^{-1}_{0,\bh{i},\bh{j}} - \hat{X}_{\bh{i} \bh{j}} - {Y_1}_{\bh{i} \bh{j}} ) ~ \GH_{\bh{j} \bh{k}} =    (\delta_{\bh{i}, \bh{j}}- \gamma_{\bh{i},\bh{j}}) \muin_{\bh{j} \bh{k}} \nn \\
&&- \left[ \hat{X}_{\bh{i} \bh{j}} \muin_{\bh{l} \bh{k} }\right] \GH_{\bh{j} \bh{m}} \mu_{\bh{m} \bh{ l}}, \label{eq65}
\earray
where the square brackets  demarcate the terms acted upon, by the  derivative operators in $\hat{X}$.
\disp{eq65} exhibits a separation of variables, all the dependence on $\mu$ is confined to the right hand side, and hence we   set both sides to the identity matrix:
\barray
&& (\GH^{-1}_{0,\bh{i},\bh{j}} - \hat{X}_{\bh{i} \bh{j}} - {Y_1}_{\bh{i} \bh{j}} ) ~ \GH_{\bh{j} \bh{k}} =\delta_{\bh{i} \bh{k}}, \label{eq66}
\earray
and
\barray
&&   (\delta_{\bh{i}, \bh{j}}- \gamma_{\bh{i},\bh{j}}) \muin_{\bh{j} \bh{k}} - \left[ \hat{X}_{\bh{i} \bh{j}} \muin_{\bh{l} \bh{k} }\right] \GH_{\bh{j} \bh{m}} \mu_{\bh{m} \bh{ l}} = \delta_{\bh{i} \bh{k}}.  \nn \\ \label{eq67}
\earray
Multiplying through with  $\mu$ and using 
\beq\left[ \hat{X}_{\bh{i} \bh{j}}  \muin_{\bh{l} \bh{k} } \right]\mu_{\bh{m} \bh{ l}}= - \left[ \hat{X}_{\bh{i} \bh{j}} \mu_{\bh{m} \bh{ l}}  \right]   \muin_{\bh{l} \bh{k} }, \label{xonmu}
\eeq
  we rewrite \disp{eq67} as 
\beq
\mu_{\bh{i}  \bh{f}} = (\delta_{\bh{i} \bh{ f} } - \gamma_{\bh{i}  \bh{f}})+ \GH_{\bh{j}  \bh{k}} \left[ \hat{X}_{\bh{i} \bh{j} } \mu_{\bh{k} \bh{f}} \right].
\eeq
We next show  that \disp{eq66} and \disp{eq67} can be rewritten in terms of the two self energies $\Phi$ and $\Psi$ used in \refdisp{ECFL} and \refdisp{Monster}. We need the relation analogous to \disp{xonmu} to simplify \disp{eq66}:
\beq
\left[ \hat{X}_{\bh{i} \bh{j}}  \GH_{\bh{j} \bh{k}}\right]=- \GH_{\bh{j} \bh{k}}\left[ \hat{X}_{\bh{i} \bh{j}}  \GHI_{ \bh{k} \bh{l} }\right]\GH_{\bh{l} \bh{k}}.
\eeq
Therefore we write the two equations as
\barray
&& (\GH^{-1}_{0,\bh{i},\bh{j}} - \Phi_{\bh{i} \bh{j}} - {Y_1}_{\bh{i} \bh{j}} ) ~ \GH_{\bh{j} \bh{k}} =\delta_{\bh{i} \bh{k}}, \label{eq662}\nn \\
&&\mu_{\bh{i}  \bh{f}} = (\delta_{\bh{i} \bh{ f} } - \gamma_{\bh{i}  \bh{f}})+ \Psi_{\bh{i} \bh{f}}, \label{exact-factors}
\earray
where the two self energies $\Phi$ and $\Psi$ are functions obtained by iteration, 
and  have a finite limit on turning off the Bosonic source $\V$. These are obtained from the above as
\barray
\Phi_{\bh{i} \bh{j}} &=& - \GH_{\bh{m} \bh{k}}\left[ \hat{X}_{\bh{i} \bh{m}}  \GHI_{ \bh{k} \bh{j} }\right] \nn \\
\Psi_{\bh{i} \bh{f}}&=&\GH_{\bh{j}  \bh{k}} \left[ \hat{X}_{\bh{i} \bh{j} } \mu_{\bh{k} \bh{f}} \right]
\earray 
Using the definition of $\hat{X}$ and of various vertex functions, we can verify that these are precisely the pair of equations that we obtained in \refdisp{ECFL} and \refdisp{Monster} for the two self energies of $\G$.


\begin{thebibliography}{99}
\bibitem{angleaverage} The thermodynamic averages  are denoted by   $\langle A \rangle \ =  \frac{1}{Z}   {\tr}_{tJ }   e^{-\beta H} A$. Here $H=H_{tJ}$ or $H=H_{eff}$ in the original description or the canonical theory description respectively.
\bibitem{comment-perturb} In most textbook examples of perturbation theory, one scales the interaction term with $\lambda$, since  typically  the interaction strength is an experimentally determined  number (e.g. the Coulomb energy in hydrogenic problems). Introducing  $\lambda$ is a device to count the  various orders of perturbation theory, and therefore a formal small parameter. 
 A true  small parameter, replacing  it in  many physical problems, is found by taking the ratio of characteristic energies of two parts of the Hamiltonian.
\bibitem{2DMFT}  In \refdisp{DMFT-ECFL} a similar exercise is carried for the large dimensional case, extending to a somewhat greater (occupied) range of $\omega$. That analysis  utilizes the  greater information available in that case.
\bibitem{calc-commute} With all operators at the same (undisplayed) site, we can show that the commutator $[C^\dagger_{\si_1} ( 1- \lambda N_{\sib_1} ), \chd{\si_2} \ch{\si_3}]$ reduces to 
$- \delta_{\si_1 \si_3} \chd{\si_2} ( 1- \lambda N_{\sib_2})$ as follows. By straightforward expansion of the commutator we get $ -   \delta_{\si_1 \si_3} \chd{\si_2} + \lambda \chd{\si_2} \left( \delta_{\si_1 \si_3} N_{\sib_1}+ \delta_{\si_2 \sib_3} (\delta_{\sib_1 \si_2}-\delta_{\sib_1 \si_3} ) \chd{\si_1} \ch{\si_3}\right)$. Here introduced the  extra term $\delta_{\si_2 \sib_3}$    using $1=\delta_{\si_2 \sib_3} +\delta_{\si_2 \si_3} $, since  the number of  components is only  two, and the vanishing of the other delta term. On using $\chd{\si_2}\chd{\si_2}=0$,  the last term then simplifies to $ \lambda  \delta_{\si_1 \si_3} \chd{\si_2} (N_{\sib_2} \delta_{\sib_1 \si_2} + N_{\sib_1})  $, and its second piece can be simplified as $ (N_{\sib_2} \delta_{\sib_1 \si_2} + N_{\sib_1}) = (N_{\si_2} \delta_{\sib_1 \si_2} + N_{\sib_2}) $. Using  $\chd{\si_2}\chd{\si_2}=0$, the required result follows.
\bibitem{wrong-count} This alternative  choice, upon using   \disp{lambda-reln2} leads to
$$
n_{i \si}= \frac{1}{\lambda} \{ 1-  \G^{(\lambda)}_{\sib \sib}( i \tau^- i \tau) +  \G^{(\lambda)}_{\sib \sib}( i \tau^+ i \tau) \} .
$$
However, this prescription does not connect continuously to the correct constitutive relation at $\lambda=1$, and is therefore rejected.
\bibitem{simplify1} Since we are interested in  $M\to \infty$ at the end,  we will simplify the notation from here by writing
 $\widetilde{c~}^*_{f \si_f}(\tau_{f+1})\to \widetilde{c~}^*_{f \si_f}(\tau_{f})$.
\bibitem{caution}
  Regarding  $V$, we need to exercise some care, since
\disp{dermaps} shows that a derivative with respect to $c_{i \si}$ is related to multiplying with $c^*$, rather than by
\bibitem{ECQL2} It is worth mentioning one possibility that was tried earlier before discussing a more practical solution  found recently in \refdisp{ECFL}.
 In \refdisp{ECQL}  a combination of two transformations is made: (1) \disp{eq60}  is left multiplied by  $(\iden - \gamma)^{-1}$ and (2) redefine $\xi \to (\iden - \gamma) ~ \xi'$. Together, these transformations give an exact formulation with a canonical Green's function $G_{i,f} = \frac{\delta \xi'_{\bh{i}} }{\delta J_{\bh{f}}}$ satisfying a canonical equation, when acted upon by  an operator $(\iden - \gamma)^{-1}~ (\GH_0 - \hat{X} - Y_1) (\iden - \gamma)$. A more thorough investigation of this scheme than in \cite{ECQL}, especially with the addition of the $\lambda$ parameter, seems to be a  worthwhile goal for the future.
\end{thebibliography}
\end{document}